\documentclass[reprint,superscriptaddress,aps,prx]{revtex4-2}
\usepackage{amsmath}
\usepackage{graphicx}
\usepackage{dcolumn}
\usepackage{subfigure}
\usepackage{dsfont}
\usepackage{hyperref}
\hypersetup{colorlinks=true, citecolor=blue, urlcolor=blue, linkcolor=blue}

\begin{document}
	\title{Quantum Advantage of Noisy Grover's Algorithm}
	
	\author{Jian Leng}
	\affiliation{ State Key Laboratory of Low Dimensional Quantum Physics, Department of Physics, \\ Tsinghua University, Beijing 100084, China}
	\author{Fan Yang}
	\affiliation{ State Key Laboratory of Low Dimensional Quantum Physics, Department of Physics, \\ Tsinghua University, Beijing 100084, China}
	\author{Xiang-Bin Wang}
	\email{ xbwang@mail.tsinghua.edu.cn}
	\affiliation{ State Key Laboratory of Low Dimensional Quantum Physics, Department of Physics, \\ Tsinghua University, Beijing 100084, China}
	\affiliation{ Jinan Institute of Quantum technology, SAICT, Jinan 250101, China}
	\affiliation{ Shanghai Branch, CAS Center for Excellence and Synergetic Innovation Center in Quantum Information and Quantum Physics, University of Science and Technology of China, Shanghai 201315, China}
	\affiliation{ Shenzhen Institute for Quantum Science and Engineering, and Physics Department, Southern University of Science and Technology, Shenzhen 518055, China}
	\affiliation{ Frontier Science Center for Quantum Information, Beijing 100193, China}
	
	\begin{abstract}
		Quantum advantage is the core of quantum computing. Grover's search algorithm is the only quantum algorithm with \textit{proven} advantage to any possible classical search algorithm. However, realizing this quantum advantage in practice is quite challenging since Grover's algorithm is very sensitive to noise. Here we present a noise-tolerant method that exponentially improves the noise threshold of Grover's algorithm. We present a lower bound for average fidelity of any quantum circuit with $\mathcal{O}(\log D\log D)$ cost under time-independent noise, where $D$ is the dimension of Hilbert space. According to this bound value, we determine the number of iterates which will be applied in Grover's algorithm. Numerical simulation shows that the noise threshold of quantum advantage of Grover's algorithm by our noise-tolerant method is improved by an exponential factor with qubit amount rise.
	\end{abstract}
	
	
	\maketitle
	
	\section{Introduction}
	Quantum advantage \cite{preskill2018quantum,aaronson2011computational,arute2019quantum,wu2021strong} plays a central role in the study of quantum computing \cite{shor1994algorithms,grover1996fast}. Though there have been lots of excellent studies \cite{google2023suppressing,zhao2022realization,zhong2021phase,pokharel2023demonstration} towards the goal, only Grover's search algorithm has been theoretically proven for quantum advantage \cite{grover1996fast}. However, realizing this quantum advantage in a practical system is technically difficult since Grover's algorithm is very sensitive to noise \cite{pablo1999robustness,shapira2003effect,shenvi2003effects,salas2008noise,reitzner2019grover,cohn2016grover}. The theoretical quantum advantage of Grover's algorithm disappears immediately  \cite{pablo1999robustness,shapira2003effect,shenvi2003effects,salas2008noise,reitzner2019grover,cohn2016grover} when the noise exceeds a small threshold in practice. Here we present a noise-tolerant method that exponentially improves the noise threshold of Grover's algorithm. In Sec. II, we explicitly construct the quantum circuit of Grover's algorithm. In Sec. III, we predict a lower bound for average fidelity of any quantum circuit with $\mathcal{O}(\log D\log D)$ cost under time-independent noise, where $D$ is the dimension of Hilbert space. We say `predict' since the lower bound is obtained before implementing the circuit. In Sec. IV, we determine the number of iterates which will be applied in Grover's algorithm. The combination of Sec. III and IV is our noise-tolerant method. In Sec. V, numerical simulation shows that the noise threshold of quantum advantage of Grover's search by our method is improved by an exponential factor with qubit amount rise.
	
	\section{Quantum circuit of Grover's algorithm}
\begin{figure}[b]
		\begin{minipage}{1\linewidth}
			\vspace{-0.2cm}\hspace{-0.9cm}\includegraphics[width=1.1\linewidth]{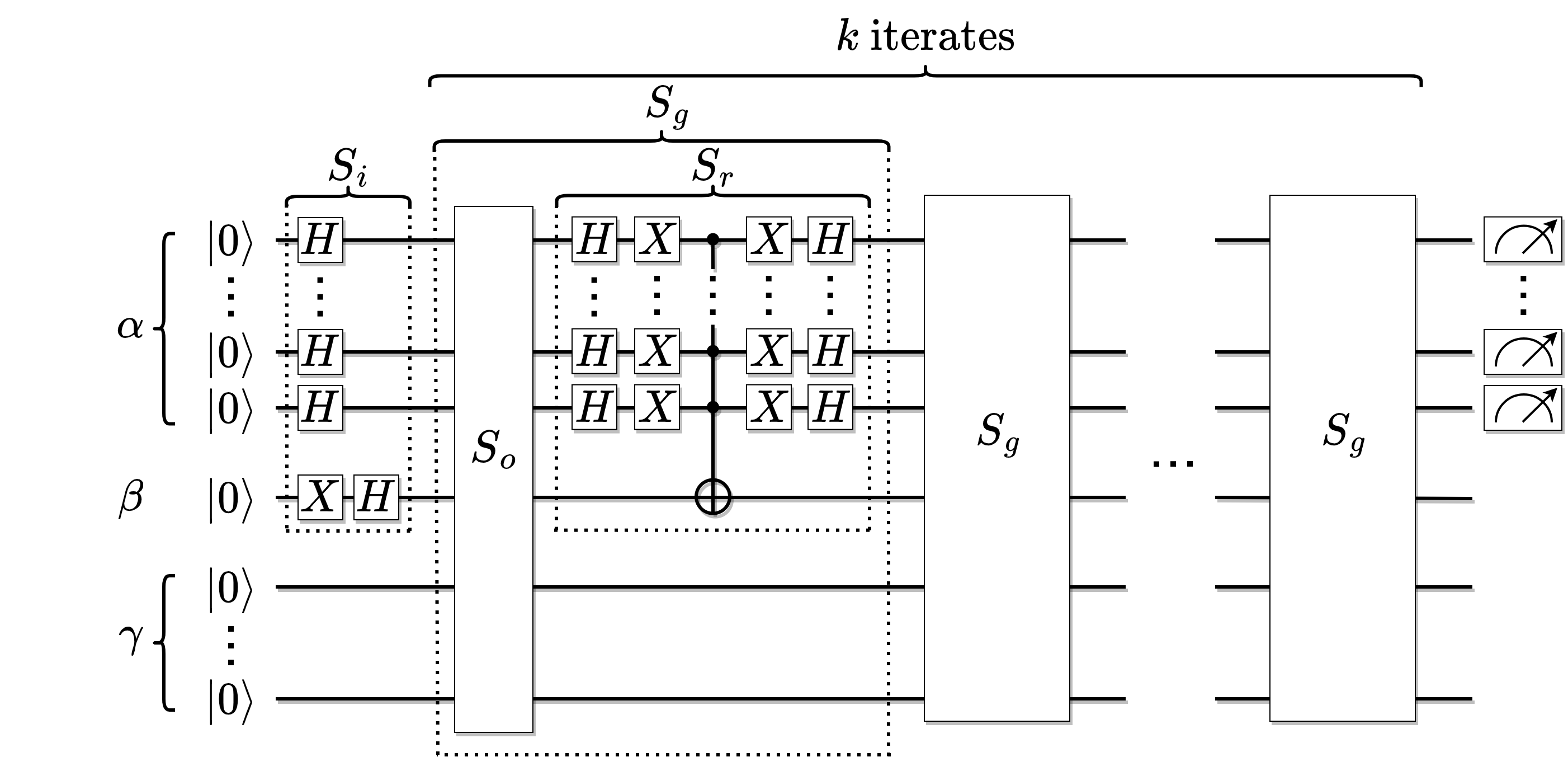}
		\end{minipage}
		\caption{\label{Grover_circuit}The quantum circuit for Grover's algorithm. Register $\alpha$, $\beta$ and $\gamma$ contain $n$, $1$ and $l$ qubits respectively. Totally $t=n+1+l$. Operator $S_i$ works on register $\alpha$ and $\beta$ initially. Oracle operator $S_o$ works on all registers and reflection operator $S_r$ works on register $\alpha$ and $\beta$. Grover operator $S_g=S_rS_o$ iterates for $k$ times. The algorithm ends with the measurement on register $\alpha$. The whole circuit is written as $U_k=S_g^kS_i$.}
	\end{figure}

	Suppose the search problem has $M$ solutions in total $N=2^n$ elements. Grover's algorithm aims to find anyone in these $M$ solutions. The circuit is shown in Fig. \ref{Grover_circuit}.
	Three registers are used: Register $\alpha$ has $n$ qubits, register $\beta$ has one qubit and register $\gamma$ has $l$ qubits. Oracle operator $S_o$ is defined as
	\begin{align}
		S_o|x\rangle_\alpha|q\rangle_\beta|0\rangle_\gamma=|x\rangle_\alpha|q+f(x)\rangle_\beta|0\rangle_\gamma,\label{oracle_operator}
	\end{align}
	where $f(x)=1$ if $x$ is a solution, otherwise $f(x)=0$. We choose a specific state for register $\beta$:
	\begin{align}
		S_o|x\rangle_\alpha\frac{|0\rangle_\beta-|1\rangle_\beta}{\sqrt{2}}|0\rangle_\gamma=(-1)^{f(x)}|x\rangle_\alpha\frac{|0\rangle_\beta-|1\rangle_\beta}{\sqrt{2}}|0\rangle_\gamma.\label{choose_beta}
	\end{align}
	Reflection operator $S_r$ is defined as
	\begin{align}
		S_r=H_\alpha^{\otimes n}X_\alpha^{\otimes n}\mathrm{Toff}_{\alpha\beta}X_\alpha^{\otimes n}H_\alpha^{\otimes n},\label{S_r}
	\end{align}
	where $\mathrm{Toff}_{\alpha\beta}$ is the $(n+1)$-qubit Toffoli gates in Fig. \ref{Grover_circuit}. We choose the same state for register $\beta$ as Eq. (\ref{choose_beta})
	\begin{align}
		&\mathrm{Toff}_{\alpha\beta}|x\rangle_\alpha\frac{|0\rangle_\beta-|1\rangle_\beta}{\sqrt{2}}\nonumber\\
		=&-(2|N-1\rangle\langle N-1|_\alpha-I_\alpha)|x\rangle_\alpha\frac{|0\rangle_\beta-|1\rangle_\beta}{\sqrt{2}}.
	\end{align}
	Then
	\begin{align}
		&S_r|x\rangle_\alpha\frac{|0\rangle_\beta-|1\rangle_\beta}{\sqrt{2}}=-H_\alpha^{\otimes n}X_\alpha^{\otimes n}\nonumber\\
		\times&(2|N-1\rangle\langle N-1|_\alpha-I_\alpha)X_\alpha^{\otimes n}H_\alpha^{\otimes n}|x\rangle_\alpha\frac{|0\rangle_\beta-|1\rangle_\beta}{\sqrt{2}}\nonumber\\
		=&-H_\alpha^{\otimes n}(2|0\rangle\langle 0|_\alpha-I_\alpha)H_\alpha^{\otimes n}|x\rangle_\alpha\frac{|0\rangle_\beta-|1\rangle_\beta}{\sqrt{2}}.\label{choose_beta_S_r}
	\end{align}
	Grover's algorithm is written as
	\begin{align}
		|\phi_f\rangle=U_k|0\rangle=S_g^kS_i|0\rangle=(S_rS_o)^k|\phi_0\rangle,\label{final_state}
	\end{align}
	where
	\begin{align}
		|\phi_0\rangle=&S_i|0\rangle=H_\alpha^{\otimes n}|0\rangle_\alpha H_\beta X_\beta|0\rangle_\beta|0\rangle_\gamma\nonumber\\
		=&\left(\sum_{x=1}^{2^n}\frac{|x\rangle_\alpha}{\sqrt{2^n}}\right)\frac{|0\rangle_\beta-|1\rangle_\beta}{\sqrt{2}}|0\rangle_\gamma.
	\end{align}
	According to Eq. (\ref{choose_beta}) and (\ref{choose_beta_S_r}), $S_o$ and $S_r$ do not change the state of register $\beta$ and $\gamma$. So we ignore the state of these two registers at this moment and simply consider $|\phi_{0v}\rangle$ in a two-dimensional space:
	\begin{align}
		|\phi_{0v}\rangle=\sqrt{1-\lambda}|R\rangle_\alpha+\sqrt{\lambda}|T\rangle_\alpha:=
		\left(
		\begin{array}{c}
			\sqrt{1-\lambda}\\
			\sqrt{\lambda}
		\end{array}
		\right),
	\end{align}
	where
	\begin{align}
		\lambda=&M/N,\nonumber\\
		|R\rangle_\alpha=&\frac{1}{\sqrt{N-M}}\sum_{x\notin\mathrm{solutions}}|x\rangle_\alpha,\nonumber\\
		|T\rangle_\alpha=&\frac{1}{\sqrt{M}}\sum_{x\in\mathrm{solutions}}|x\rangle_\alpha.\label{target_state}
	\end{align}
	The oracle, reflection and Grover operators can also be written in this two-dimensional space:
	\begin{align}
		S_{ov}=&\left(
		\begin{array}{cc}
			1 & 0\\
			0 & -1
		\end{array}
		\right),\nonumber\\
		S_{rv}=&-\left(
		\begin{array}{cc}
			1-2\lambda &  2\sqrt{\lambda-\lambda^2} \\
			2\sqrt{\lambda-\lambda^2} & 2\lambda-1
		\end{array}
		\right),\nonumber\\
		S_{gv}=&S_{rv}S_{ov}=-\left(
		\begin{array}{cc}
			1-2\lambda &  -2\sqrt{\lambda-\lambda^2} \\
			2\sqrt{\lambda-\lambda^2} & 1-2\lambda
		\end{array}
		\right).
	\end{align}
	The diagonalized form of $S_{gv}$ is
	\begin{align}
		S_{gv}=-V\Lambda V^{-1},
	\end{align}
	where
	\begin{align}
		&V=\frac{1}{\sqrt{2}}\left(
		\begin{array}{cc}
			1 &  1 \\
			-i & i
		\end{array}
		\right),&\Lambda=\left(
		\begin{array}{cc}
			e^{i\theta}&0 \\
			0&e^{-i\theta}
		\end{array}
		\right),
	\end{align}
	$\sin(\theta/2)=\sqrt{\lambda}$ and $\cos(\theta/2)=\sqrt{1-\lambda}$. So $S_{gv}^k=(-1)^kV\Lambda^k V^{-1}$ and we obtain the final state in this two-dimensional space from Eq. (\ref{final_state}):
	\begin{align}
		(-1)^k|\phi_f\rangle=&\cos\left[(k+\frac{1}{2})\theta\right]|R\rangle_\alpha+\sin\left[(k+\frac{1}{2})\theta\right]|T\rangle_\alpha\nonumber\\
		&\otimes\frac{|0\rangle_\beta-|1\rangle_\beta}{\sqrt{2}}|0\rangle_\gamma.\label{final_state_two_dimension}
	\end{align}
	Measuring $|\phi_f\rangle$ in register $\alpha$ we obtain a correct result with success probability
	\begin{align}
		s(k)=\sin^2\left[(2k+1)\theta/2\right].\label{success_probability_noiseless}
	\end{align}
	The original Grover's algorithm chooses 
	\begin{align}
		k=k_g=\lfloor\pi/(2\theta)-1/2\rceil\label{kg}
	\end{align}
	This gives the optimal success probability $s(k_g)$ that converges to $100\%$ when $\lambda\to0$. But the tight bound for scale of running time is obtained by minimizing $r(k)=k/s(k)$ \cite{boyer1998tight}. Applying $\mathrm{d}r(k)/\mathrm{d}k=0$ we have 
	\begin{align}
		2k\theta=\tan(k\theta+\frac{\theta}{2}).\label{kt}
	\end{align}
	Numerically solving this equation we obtain a solution $k_t$. Then choosing $k=\lfloor k_t\rceil$ we get the optimal running time $r(\lfloor k_t\rceil)$ that converges to $0.88\times r(k_g)$ when $\lambda\to0$.
	
	\section{Predicting lower bound for average fidelity}
	
	We call a single-qubit gate or a two-qubit gate for two neighboring qubits a \textit{local gate}. Practically, any quantum circuit is supposed to be decomposed into local gates that can be implemented in experiment. We use the notation $G^{(i)}_a$ to represent a local gate $G^{(i)}$ applied on one or two qubits that are denoted by $a$. For example, $X_5$ applies a Pauli-X gate on the 5th qubit and $\mathrm{CNOT}_{7,8}$ applies a controlled-NOT gate on 7th and 8th qubits. Explicitly, in a $t$-qubit system, we use
	\begin{align}
		\tilde{G}^{(i)}_a:=G^{(i)}_a\otimes I_{\bar{a}}
	\end{align}
	for local gate $G^{(i)}_a$, where $\bar{a}$ denotes all qubits except $a$.
	
	Part 1 of our noise-tolerant method: We decompose an arbitrary quantum circuit $U_{\mathrm{circuit}}$ by local gates set $\{\tilde{G}^{(i)}_a\}$ that contains $\mathcal{O}(t)$ elements:
	\begin{equation}
		U_{\mathrm{circuit}}=...\tilde{G}^{(k)}_c\tilde{G}^{(j)}_b\tilde{G}^{(i)}_a.\label{U_circuit}
	\end{equation}
	This decomposition can always be done, since any quantum circuit $U_{\mathrm{circuit}}$ can be decomposed with universal quantum gates set \cite{boykin1999universal}:
	\begin{align}
		\{&\widetilde{\mathrm{CNOT}}_{1,2},~\widetilde{\mathrm{CNOT}}_{2,3},~...,~\widetilde{\mathrm{CNOT}}_{t-1,t};~\tilde{T_1},~\tilde{T_2},~...,~\tilde{T_t};\nonumber\\
		&\tilde{H_1},~\tilde{H_2},~...,~\tilde{H_t};~\tilde{S_1},~\tilde{S_2},~...,~\tilde{S_t}\}
	\end{align}
	which is a subset of local gates set $\{\tilde{G}^{(i)}_a\}$. Remember the notation $\sim$ means that $\widetilde{\mathrm{CNOT}}_{1,2}=\mathrm{CNOT}_{1,2}\otimes I_{\overline{1,2}},~\tilde{T_1}=T_1\otimes I_{\bar{1}}$ and so on.	However, instead of ideal local gate $\tilde{G}^{(i)}_a$, we can only implement noisy quantum channel $\tilde{\mathcal{G}}^{(i)}_a$ in practical environment. Suppose the noise is time-independent, i.e., implementing $\tilde{G}^{(i)}_a$ at different time corresponds to the same noisy quantum channel $\tilde{\mathcal{G}}^{(i)}_a$. The noise is quite general that it could be global depolarizing, local bit-flip or the mixture of several different types of noise. It is not limited to the individual noise. The noise on $\tilde{G}^{(i)}_a$ could be different from the noise on another local gate $\tilde{G}^{(j)}_b$.
	
	\textit{Fidelity} represents how close between two density matrices $\rho$ and $\sigma$:
	\begin{equation}
		F(\rho,\sigma)=\mathrm{tr}\left(\sqrt{\rho^{1/2}\sigma\rho^{1/2}}\right)^2.\label{Fidelity}
	\end{equation}
	Especially, when $\sigma$ is a pure state $|\psi\rangle$, it becomes to $F(\rho,|\psi\rangle)=\langle\psi|\rho|\psi\rangle$.
	
	\textit{Jamiolkowski isomorphism} \cite{jamiolkowski1972linear} maps a unitary operation $G$ to a pure state $|\psi_G\rangle$ or maps a quantum channel $\mathcal{G}$ to a density matrix $\rho_{\tilde{G}}$:
	\begin{align}
		|\psi_G\rangle=&(G\otimes I)|\Omega\rangle,\nonumber\\
		\rho_{\mathcal{G}}=&(\mathcal{G}\otimes I)(|\Omega\rangle\langle\Omega|),\label{Jamiolkowski isomorphism}
	\end{align}
	where $|\Omega\rangle=\sum_k|k\rangle|k\rangle/\sqrt{D}$ is a maximally entangled state of original $D$-dimensional space with a copy of itself. 
	
	\textit{Process fidelity} represents how close between ideally unitary gate $G$ and noisy quantum channel $\mathcal{G}$:
	\begin{equation}
		F_{\mathrm{pro}}(G,\mathcal{G}):=F(|\psi_G\rangle,\rho_{\mathcal{G}})=\langle\psi_G|\rho_{\mathcal{G}}|\psi_G\rangle.\label{F_pro}
	\end{equation}
	
	Part 2 of our noise-tolerant method: We apply local gates set $\{\tilde{G}^{(i)}_a\}$ and measure its local process fidelity. This includes three steps:
	
	1) Prepare initial state $\rho_a\otimes \mathds{1}_{\bar{a}}$. Apply $\tilde{G}^{(i)}_a$ and obtain $\tilde{\mathcal{G}}^{(i)}_a(\rho_a\otimes \mathds{1}_{\bar{a}})$. Apply quantum process tomography or other technologies on $\rho_a$ to get the local process fidelity $F_{\mathrm{pro}}(G^{(i)}_a,\mathcal{G}^{(i)}_a)$ where $\mathcal{G}^{(i)}_a(\rho_a):=\mathrm{tr}_{\bar{a}}[\tilde{\mathcal{G}}^{(i)}_a(\rho_a\otimes \mathds{1}_{\bar{a}})]$.
	
	2) Prepare initial state $\rho_b\otimes \mathds{1}_{\bar{b}}$ where $b$ represents single qubit and $b\notin a$. Apply $\tilde{G}^{(i)}_a$ to get $\tilde{\mathcal{G}}^{(i)}_a(\rho_b\otimes \mathds{1}_{\bar{b}})$. Apply quantum process tomography or other technologies on $\rho_b$ to get the local process fidelity $F_{\mathrm{pro}}(I_b,\mathcal{I}^{(i)}_{ab})$ where $\mathcal{I}^{(i)}_{ab}(\rho_b):=\mathrm{tr}_{\bar{b}}[\tilde{\mathcal{G}}^{(i)}_a(\rho_b\otimes \mathds{1}_{\bar{b}})]$. Repeat this step for every $b$ except the single or two qubits denoted by $a$.
	
	3) Repeat the first two steps for every $\tilde{G}^{(i)}_a$.
	
	The sum of cost of step 1) and 2) is $\mathcal{O}(t)$. These two steps repeat $\mathcal{O}(t)$ times in step 3) since local gates set $\{\tilde{G}^{(i)}_a\}$ contains $\mathcal{O}(t)$ elements. So the total cost of these three steps is $\mathcal{O}(t^2)$ or $\mathcal{O}(\log D\log D)$ where $D=2^t$ is the dimension of $t$-qubit Hilbert space. \textit{Local estimator} for $\tilde{G}^{(i)}_a$ is defined as \cite{dive2017controlling,dive2018situ}
	\begin{align}
		&F_{LE}(\tilde{G}^{(i)}_a):=\nonumber\\
		&\max\left[F_{\mathrm{pro}}(G^{(i)}_a,\mathcal{G}^{(i)}_a)-\sum_{b\ne a}\left(1-F_{\mathrm{pro}}(I_b,\mathcal{I}^{(i)}_{ab})\right),0\right].\label{local_estimator}
	\end{align}
	It is proved that a lower bound for process fidelity $\tilde{G}^{(i)}_a$ is given by \cite{dive2017controlling,dive2018situ}:
	\begin{equation}
		F_{\mathrm{pro}}(\tilde{G}^{(i)}_a,\tilde{\mathcal{G}}^{(i)}_a)\ge F_{LE}(\tilde{G}^{(i)}_a).\label{U_a_process_fidelity}
	\end{equation}
	Process fidelity $F_{\mathrm{pro}}$ can be turned into three metrics:
	$\sqrt{2-2\sqrt{F_{\mathrm{pro}}}}$, $\arccos\sqrt{F_{\mathrm{pro}}}$ and $\sqrt{1-F_{\mathrm{pro}}}$ \cite{gilchrist2005distance}. The \textit{chaining property} of the first metric gives
	\begin{align}
		&\sqrt{2-2\sqrt{F_{\mathrm{pro}}(\tilde{G}^{(j)}_b\tilde{G}^{(i)}_a,\tilde{\mathcal{G}}^{(j)}_b\tilde{\mathcal{G}}^{(i)}_a)}}\nonumber\\
		\le&\sqrt{2-2\sqrt{F_{\mathrm{pro}}(\tilde{G}^{(j)}_b,\tilde{\mathcal{G}}^{(j)}_b)}}+\sqrt{2-2\sqrt{F_{\mathrm{pro}}(\tilde{G}^{(i)}_a,\tilde{\mathcal{G}}^{(i)}_a)}}\nonumber\\
		\le&\sqrt{2-2\sqrt{F_{LE}(\tilde{G}^{(j)}_b)}}+\sqrt{2-2\sqrt{F_{LE}(\tilde{G}^{(i)}_a)}},
	\end{align}
	where Eq. (\ref{U_a_process_fidelity}) has been used for the second inequality. Then we have
	\begin{align}
		&F_{\mathrm{pro}}(\tilde{G}^{(j)}_b\tilde{G}^{(i)}_a,\tilde{\mathcal{G}}^{(j)}_b\tilde{\mathcal{G}}^{(i)}_a)\nonumber\\ \ge&g_1\left(F_{\mathrm{pro}}(\tilde{G}^{(j)}_b,\tilde{\mathcal{G}}^{(j)}_b),F_{\mathrm{pro}}(\tilde{G}^{(i)}_a,\tilde{\mathcal{G}}^{(i)}_a)\right)\nonumber\\
		\ge&g_1\left(F_{LE}(\tilde{G}^{(j)}_b),F_{LE}(\tilde{G}^{(i)}_a)\right),
	\end{align}
	where
	\begin{align}
		g_1(x,y)=\max\left[0,1-\left(\sqrt{1-\sqrt{x}}+\sqrt{1-\sqrt{y}}\right)^2\right]^2\label{g1}
	\end{align}
	with $\max(x,y)=x$ if $x\ge y$, otherwise $\max(x,y)=y$. Similar results arise for the last two metrics:
	\begin{align}
		&F_{\mathrm{pro}}(\tilde{G}^{(j)}_b\tilde{G}^{(i)}_a,\tilde{\mathcal{G}}^{(j)}_b\tilde{\mathcal{G}}^{(i)}_a)\nonumber\\ \ge&g_2\left(F_{\mathrm{pro}}(\tilde{G}^{(j)}_b,\tilde{\mathcal{G}}^{(j)}_b),F_{\mathrm{pro}}(\tilde{G}^{(i)}_a,\tilde{\mathcal{G}}^{(i)}_a)\right)\nonumber\\
		\ge&g_2\left(F_{LE}(\tilde{G}^{(j)}_b),F_{LE}(\tilde{G}^{(i)}_a)\right),\nonumber\\
		&F_{\mathrm{pro}}(\tilde{G}^{(j)}_b\tilde{G}^{(i)}_a,\tilde{\mathcal{G}}^{(j)}_b\tilde{\mathcal{G}}^{(i)}_a)\nonumber\\ \ge&g_3\left(F_{\mathrm{pro}}(\tilde{G}^{(j)}_b,\tilde{\mathcal{G}}^{(j)}_b),F_{\mathrm{pro}}(\tilde{G}^{(i)}_a,\tilde{\mathcal{G}}^{(i)}_a)\right)\nonumber\\
		\ge&g_3\left(F_{LE}(\tilde{G}^{(j)}_b),F_{LE}(\tilde{G}^{(i)}_a)\right),
	\end{align}
	where
	\begin{align}
		&g_2(x,y)=\max\left[0,\cos(\arccos\sqrt{x}+\arccos\sqrt{y})\right]^2,\nonumber\\
		&g_3(x,y)=1-\left(\sqrt{1-x}+\sqrt{1-y}\right)^2.\label{g2_g3}
	\end{align}
	So the lower bound for process fidelity of two gates $\tilde{G}^{(j)}_b\tilde{G}^{(i)}_a$ is
	\begin{align}
		&F_{\mathrm{pro}}(\tilde{G}^{(j)}_b\tilde{G}^{(i)}_a,\tilde{\mathcal{G}}^{(j)}_b\tilde{\mathcal{G}}^{(i)}_a)\nonumber\\ \ge&g\left(F_{\mathrm{pro}}(\tilde{G}^{(j)}_b,\tilde{\mathcal{G}}^{(j)}_b),F_{\mathrm{pro}}(\tilde{G}^{(i)}_a,\tilde{\mathcal{G}}^{(i)}_a)\right)\nonumber\\
		\ge&g\left(F_{LE}(\tilde{G}^{(j)}_b),F_{LE}(\tilde{G}^{(i)}_a)\right),\label{two_gates_process_fidelity}
	\end{align}
	where
	\begin{align}
		g(x,y)=\max\left[g_1(x,y),~g_2(x,y),~g_3(x,y)\right].\label{g}
	\end{align}
	Similarly, we get a lower process fidelity bound for product of three gates $\tilde{G}^{(k)}_c\tilde{G}^{(j)}_b\tilde{G}^{(i)}_a$ by dividing it into $\tilde{G}^{(k)}_c$ and $\tilde{G}^{(j)}_b\tilde{G}^{(i)}_a$:
	\begin{align}
		&F_{\mathrm{pro}}(\tilde{G}^{(k)}_c\cdot \tilde{G}^{(j)}_b\tilde{G}^{(i)}_a,~\tilde{\mathcal{G}}^{(k)}_c\cdot\tilde{\mathcal{G}}^{(j)}_b\tilde{\mathcal{G}}^{(i)}_a)\nonumber\\
		\ge&g\left[F_{\mathrm{pro}}(\tilde{G}^{(k)}_c,\tilde{\mathcal{G}}^{(k)}_c),F_{\mathrm{pro}}(\tilde{G}^{(j)}_b\tilde{G}^{(i)}_a,\tilde{\mathcal{G}}^{(j)}_b\tilde{\mathcal{G}}^{(i)}_a)\right]\nonumber\\
		\ge &g\left[F_{LE}(\tilde{G}^{(k)}_c),~g\left(F_{LE}(\tilde{G}^{(j)}_b),F_{LE}(\tilde{G}^{(i)}_a)\right)\right].\label{three_gates_process_fidelity}
	\end{align}
	
	Part 3 of our noise-tolerant method: We calculate for the product of more gates in Eq. (\ref{U_circuit}) and finally obtain a lower bound for process fidelity of quantum circuit $U_{\mathrm{circuit}}$:
	\begin{align}
		&F_{\mathrm{pro}}(U_{\mathrm{circuit}},\mathcal{U}_{\mathrm{circuit}})\ge\nonumber\\
		&g\left\{...,~g\left[F_{\mathrm{pro}}(\tilde{G}^{(k)}_c,\tilde{\mathcal{G}}^{(k)}_c),F_{\mathrm{pro}}(\tilde{G}^{(j)}_b\tilde{G}^{(i)}_a,\tilde{\mathcal{G}}^{(j)}_b\tilde{\mathcal{G}}^{(i)}_a)\right]\right\}\ge\nonumber\\
		 &g\left\{...,~g\left[F_{LE}(\tilde{G}^{(k)}_c),~g\left(F_{LE}(\tilde{G}^{(j)}_b),F_{LE}(\tilde{G}^{(i)}_a)\right)\right]\right\}.\label{U_circuit_process_fidelity}
	\end{align}
	
	\textit{Average fidelity} represents the difference between the output states from ideally unitary gate $G$ and noisy quantum channel $\mathcal{G}$ where all input states $|\psi\rangle$ are averaged:
	\begin{equation}
		F_{\mathrm{ave}}(G,\mathcal{G})=\int d\psi F\left(G|\psi\rangle,\mathcal{G}(\psi)\right).
	\end{equation}
	The lower bound for average fidelity of $U_{\mathrm{circuit}}$ can be obtained by applying the process fidelity in Eq. (\ref{U_circuit_process_fidelity}) \cite{horodecki1999general,nielsen2002simple}:
	\begin{align}
		F_{\mathrm{ave}}(U_{\mathrm{circuit}},\mathcal{U}_{\mathrm{circuit}})=\frac{F_{\mathrm{pro}}(U_{\mathrm{circuit}},\mathcal{U}_{\mathrm{circuit}})D+1}{D+1},\label{U_circuit_average_fidelity}
	\end{align}
	where $D=2^t$ is the dimension of $t$-qubit Hilbert space.
	
	Part 4 of our noise-tolerant method: Applying the lower bound Eq. (\ref{U_circuit_average_fidelity}), we calculate the iterates number $k_0$ for Grover's algorithm by the method shown in Sec. IV. We carry out Grover's algorithm with $k_0$ iterates.
	
	Our noise-tolerant method is summarized into four parts: Firstly, decompose the $t$-qubit quantum circuit $U_{\mathrm{circuit}}$ with the local gates set $\{\tilde{G}^{(i)}_a\}$ that contains $\mathcal{O}(t)$ elements as shown in Eq. (\ref{U_circuit}). This makes part 1 of our method. Secondly, get local estimators for every local gate in $\{\tilde{G}^{(i)}_a\}$ from experiment. The cost is $\mathcal{O}(\log D\log D)$. Detailed procedure is presented in the content between Eq. (\ref{F_pro}) and Eq. (\ref{local_estimator}). This makes part 2 of our method. Thirdly, calculate the lower bound for average fidelity of $U$ by applying Eq. (\ref{U_circuit_process_fidelity}) and (\ref{U_circuit_average_fidelity}). This does not cost resource since it can be efficiently solved on classical computer. This makes part 3 of our method. Finally, calculate $k_0$ and carry out Grover's algorithm with $k_0$ iterates. This makes part 4 of our method as shown in Sec. IV.
	
	\section{Calculation method for $k_0$}
	
	Suppose that we have obtained local estimators Eq. (\ref{local_estimator}) for every local gate $\tilde{G}^{(i)}_a$ from experiment. So we can calculate the lower bound for process fidelity of $S_i$ and $S_g$ by applying Eq. (\ref{U_circuit_process_fidelity}):
	\begin{align}
		F_{\mathrm{pro}}(S_i,\mathcal{S}_i)\ge C_i,~~F_{\mathrm{pro}}(S_g,\mathcal{S}_g)\ge C_g.\label{Si_Sg_process_fidelity}
	\end{align}
	Applying Eq. (\ref{two_gates_process_fidelity}--\ref{U_circuit_process_fidelity}) we have
	\begin{align}
		F_{\mathrm{pro}}(S_g^2,\mathcal{S}_g^2)\ge&g(C_g,C_g),\nonumber\\
		F_{\mathrm{pro}}(S_g^3,\mathcal{S}_g^3)\ge&g[C_g,~g(C_g,C_g)],\nonumber\\
		...&\nonumber\\
		F_{\mathrm{pro}}(S_g^k,\mathcal{S}_g^k)\ge& g\{...,~g[C_g,~g(C_g,C_g)]\}:=C(k),\nonumber\\
		F_{\mathrm{pro}}(U_k,\mathcal{U}_k)=&F_{\mathrm{pro}}(S_g^kS_i,\mathcal{S}_g^k\mathcal{S}_i)\ge g(C(k),C_i),
	\end{align}
	where $U_k=S_g^kS_i$ is the Grover circuit in Fig. \ref{Grover_circuit}. Applying Eq. (\ref{U_circuit_average_fidelity}) we obtain the lower bound for average fidelity of $U_k$:
	\begin{align}
		F_{\mathrm{ave}}(U_k,\mathcal{U}_k)\ge&\frac{g\left(C(k),C_i\right)D+1}{D+1}:=p(k).\label{Grover_average_fidelity}
	\end{align}
	The final state of Grover's algorithm in noisy environment is $\mathcal{U}_k(|0\rangle\langle0|)$. We measure the register $\alpha$ and obtain a correct result with the following success probability
	\begin{align}
		s^\prime(k)=F[|T\rangle_\alpha,\mathrm{tr}_\beta(\mathcal{U}_k(|0\rangle\langle0|))].
	\end{align}
	According to Ref. \cite{barnum1996noncommuting}:
	\begin{align}
		F[\mathrm{tr}_\beta(\rho),\mathrm{tr}_\beta(\sigma)]\ge F[\rho,\sigma].
	\end{align}
	So we have 
	\begin{align}
		s^\prime(k)\ge \max_\psi F[|T\rangle_\alpha\otimes|\psi\rangle_\beta,\mathcal{U}_k(|0\rangle\langle0|)].
	\end{align}
	Applying \textit{triangle inequality} of the metric $\sqrt{2-2\sqrt{F}}$ we obtain
	\begin{align}
		&\sqrt{2-2\sqrt{s^\prime(k)}}\le\sqrt{2-2\sqrt{\max_\psi F(|T\rangle_\alpha\otimes|\psi\rangle_\beta,U_k|0\rangle)}}\nonumber\\
		+&\sqrt{2-2\sqrt{F[U_k|0\rangle,\mathcal{U}_k(|0\rangle\langle0|)]}}.\label{triangle_inequality}
	\end{align}
	Using Eqs. (\ref{final_state}), (\ref{final_state_two_dimension}) and (\ref{success_probability_noiseless}) to the first term of right side of Eq. (\ref{triangle_inequality}) we have
	\begin{align}
		\max_\psi F(|T\rangle_\alpha\otimes|\psi\rangle_\beta,U_k|0\rangle)=&\max_\psi F(|T\rangle_\alpha\otimes|\psi\rangle_\beta,|\phi_f\rangle)\nonumber\\
		=s(k)=&\sin^2\left[(2k+1)\theta/2\right].
	\end{align}
	For the second term there we make an approximation
	\begin{align}
		F[U_k|0\rangle,\mathcal{U}_k(|0\rangle\langle0|)]\approx&\int d\psi F[U_k|\psi\rangle,\mathcal{U}_k(|\psi\rangle\langle\psi|)]\nonumber\\
		=&F_{\mathrm{ave}}(U_k,\mathcal{U}_k)\ge p(k),
	\end{align}
	where Eq. (\ref{Grover_average_fidelity}) has been used. Now Eq. (\ref{triangle_inequality}) gives a approximate lower bound for success probability
	\begin{align}
		s^\prime(k)\gtrsim g_1(s(k),p(k)).
	\end{align}
	where $g_1(x,y)$ is given by Eq. (\ref{g1}). Similarly, applying the other two metrics $\arccos\sqrt{F}$ and $\sqrt{1-F}$ we obtain
	\begin{align}
		s^\prime(k)\gtrsim g_2(s(k),p(k));~~~s^\prime(k)\gtrsim g_3(s(k),p(k)).
	\end{align}
	With Eq. (\ref{g}), we obtain the following approximate lower bound for success probability and upper bound for running time
	\begin{align}
		s^\prime(k)\gtrsim& g(s(k),p(k)),\nonumber\\
		r^\prime(k)=&\frac{k}{s^\prime(k)}\lesssim\frac{k}{g(s(k),p(k))}:=h(k).\label{success_probability_running time}
	\end{align}
	Minimizing the approximate upper running time bound $h(k)$ we obtain the solution $k=k_0$. We carry out Grover's algorithm with $k_0$ iterates instead of $k_g$ or $\lfloor k_t\rceil$ iterates in Fig. \ref{Grover_circuit}. If the approximate upper running time bound $h(k_0)$ is higher than the classically random sampling running time $\frac{N}{2M}$, we take $k_0=1$, since more iterates can only produce worse results.
	
	\section{Numerical simulation}
	We shall numerically calculate the running time of our noise-tolerant method above and compare it with original Grover' algorithm and tight bound of Ref. \cite{boyer1998tight}. We use the scale of running time \cite{boyer1998tight}
	\begin{align}
		r=k/s
	\end{align}
	where $k$ is the iterates number and $s$ is the success probability. The original Grover's algorithm, tight bound of Ref. \cite{boyer1998tight} and our noise-tolerant method choose $k=k_g$ (Eq. (\ref{kg})), $k=\lfloor k_t\rceil$ (Eq. (\ref{kt})) and $k=k_0$ (Eq. (\ref{success_probability_running time})) respectively. Numerical simulating the Grover circuit with $k_g$, $k_t$ and $k_0$ iterates we obtain the output density matrices $\rho_g$, $\rho_t$ and $\rho_0$. Measuring these output density matrices in register $\alpha$ we get the success probability $s=s_g$, $s=s_t$ and $s=s_0$ where each $s$ is the fidelity of $\mathrm{tr}_\beta(\rho)$ and target state $|T\rangle_\alpha$ in Eq. (\ref{target_state}). Explicitly,
	\begin{align}
		s_g =& _\alpha\langle T|\mathrm{tr}_\beta(\rho_g)|T\rangle_\alpha,\nonumber\\
		s_t =& _\alpha\langle T|\mathrm{tr}_\beta(\rho_t)|T\rangle_\alpha,\nonumber\\
		s_0 =& _\alpha\langle T|\mathrm{tr}_\beta(\rho_0)|T\rangle_\alpha.
	\end{align}
	
\begin{figure}[htbp]
		\begin{minipage}{0.45\linewidth}
			\vspace{0.1cm}
			\hspace{-1.3cm}\includegraphics[width=1\linewidth]{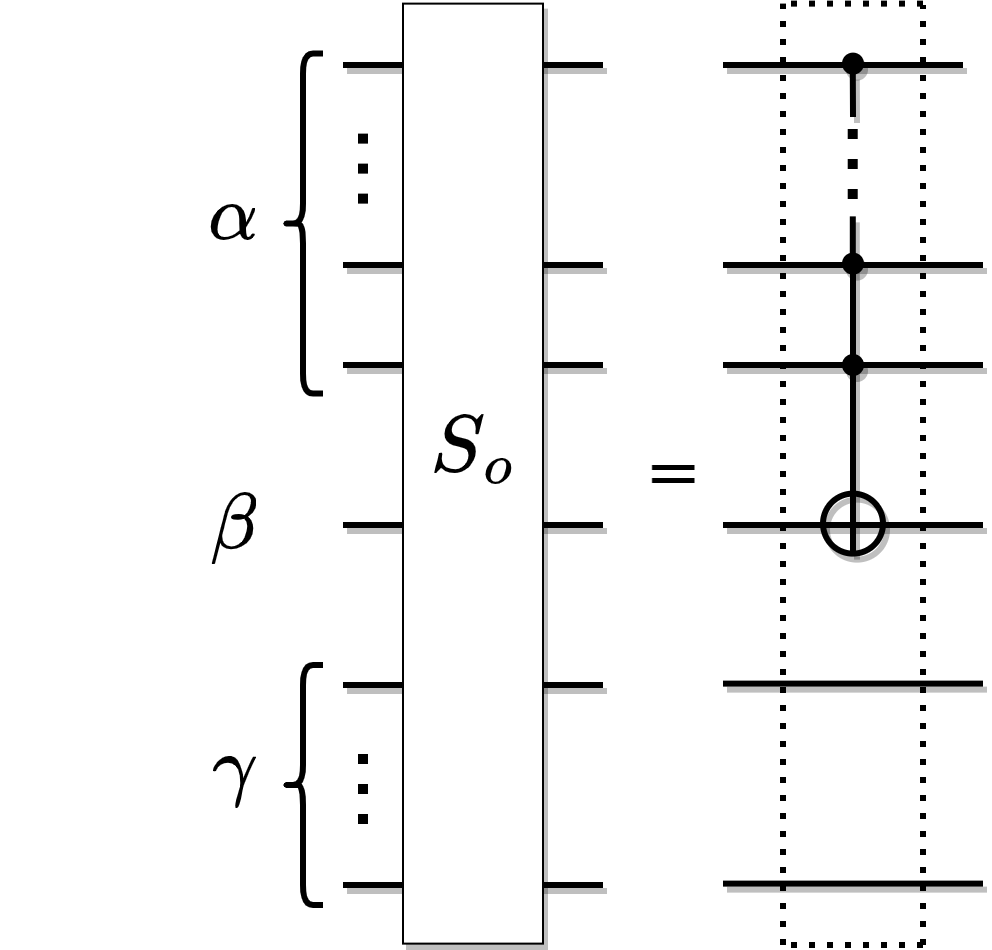}
		\end{minipage}
	
		\caption{\label{oracle_circuit}The oracle choice in this paper. We have dropped register $\gamma$ in Grover circuit since it does not work in this case.}
	\end{figure}

\begin{figure*}
		\leftline{\hspace{-1.7cm}(a) Bit-flip noise.}
		
		\leftline{\hspace{-1.1cm}$n=5$, $t=6$.~~~~~~~~~~~~~~~~~~~~~~~~~~~~~~~$n=6$, $t=7$.~~~~~~~~~~~~~~~~~~~~~~~~~~~~~~~~$n=7$, $t=8$.~~~~~~~~~~~~~~~~~~~~~~~~~~~~~~~$n=8$, $t=9$.}
		
		\hspace{-2.4cm}
		\begin{minipage}{0.245\linewidth}
			\includegraphics[width=1.3\linewidth]{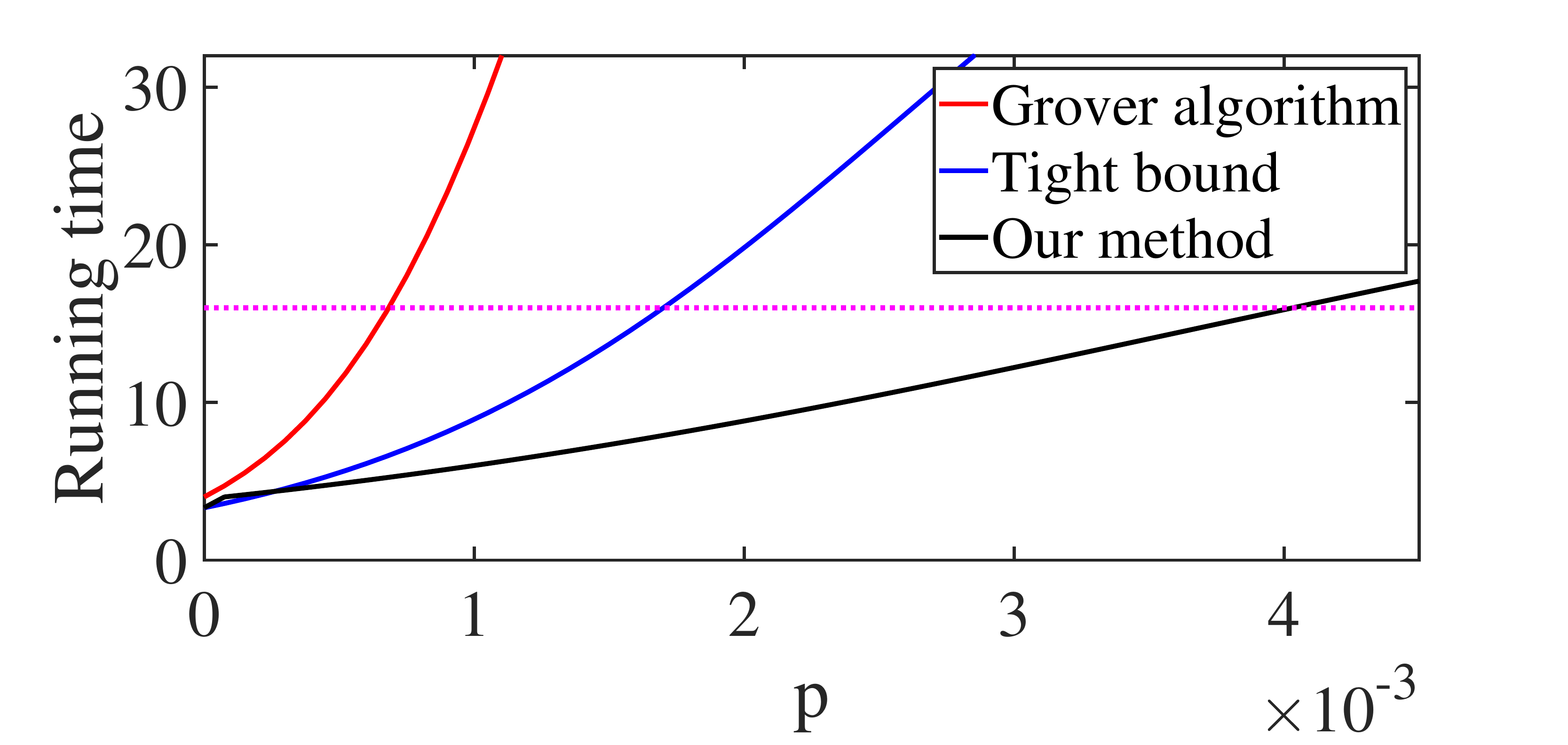}
		\end{minipage}
		\hspace{0.6cm}
		\begin{minipage}{0.245\linewidth}
			\includegraphics[width=1.3\linewidth]{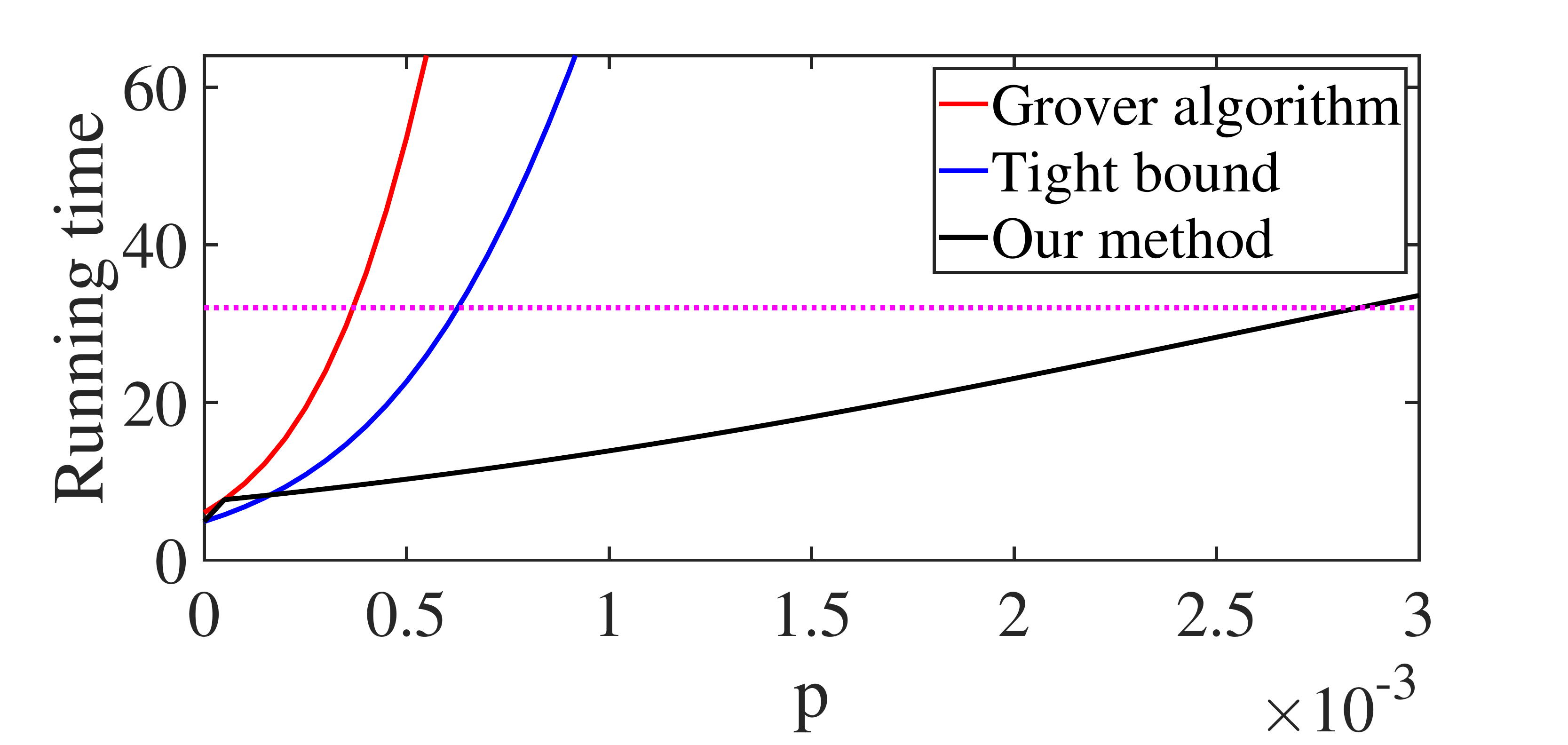}
		\end{minipage}
		\hspace{0.6cm}
		\begin{minipage}{0.245\linewidth}
			\includegraphics[width=1.3\linewidth]{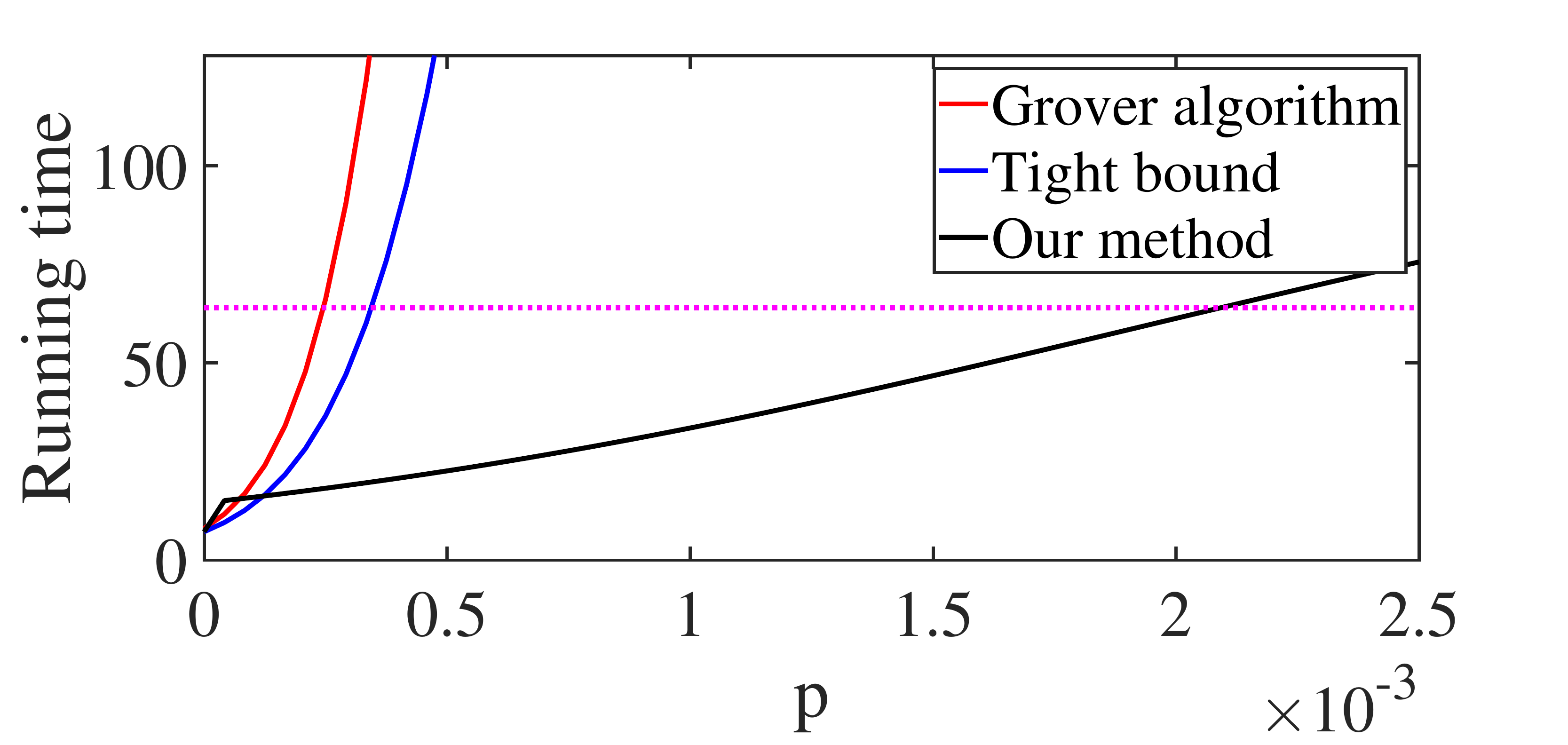}
		\end{minipage}
		\hspace{0.6cm}
		\begin{minipage}{0.245\linewidth}
			\includegraphics[width=1.3\linewidth]{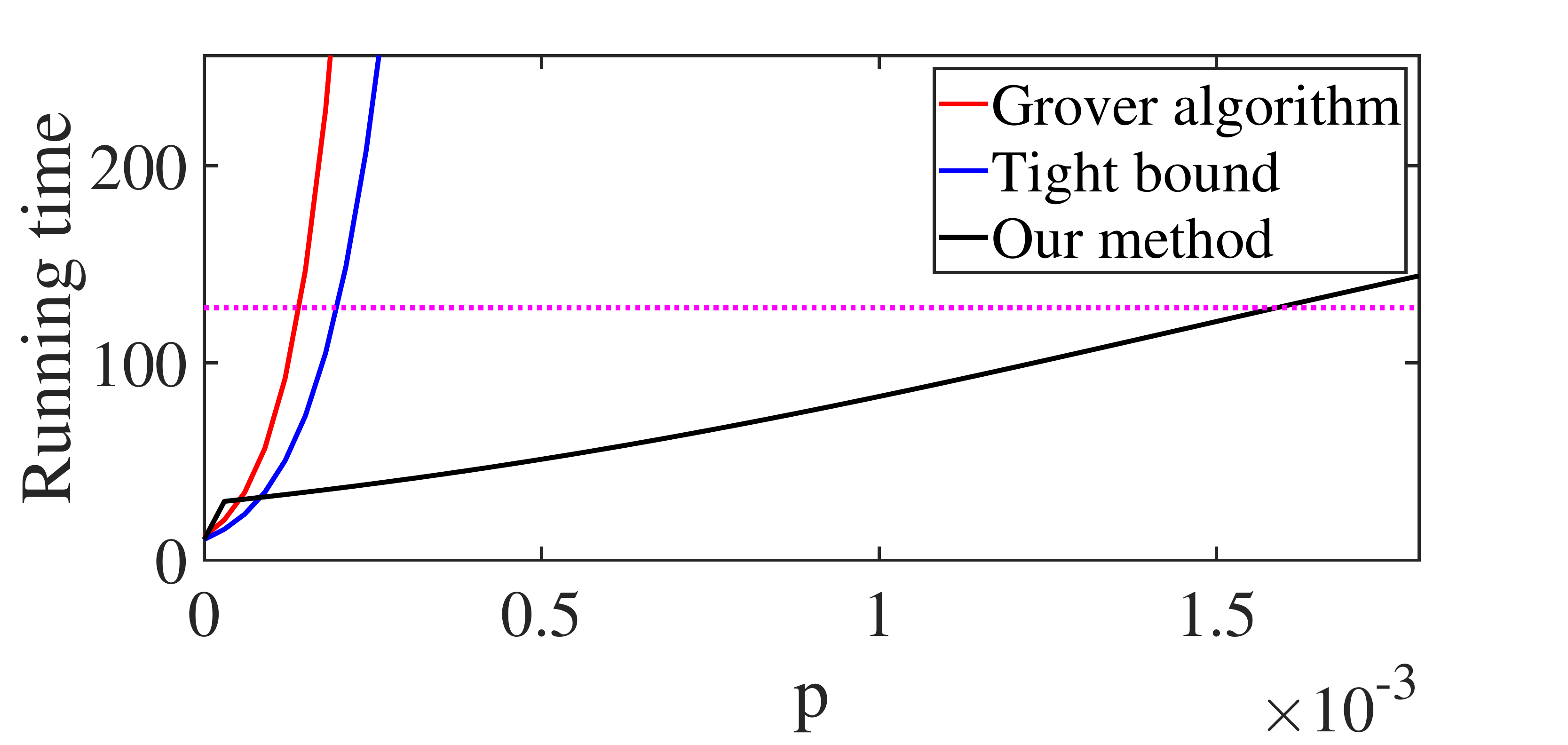}
		\end{minipage}
		\hspace{-1cm}
		
		\leftline{\hspace{-1.7cm}(b) Phase-flip noise.}
		
		\leftline{\hspace{-1.1cm}$n=5$, $t=6$.~~~~~~~~~~~~~~~~~~~~~~~~~~~~~~~$n=6$, $t=7$.~~~~~~~~~~~~~~~~~~~~~~~~~~~~~~~~$n=7$, $t=8$.~~~~~~~~~~~~~~~~~~~~~~~~~~~~~~~$n=8$, $t=9$.}
		
		\hspace{-2.4cm}
		\begin{minipage}{0.245\linewidth}
			\includegraphics[width=1.3\linewidth]{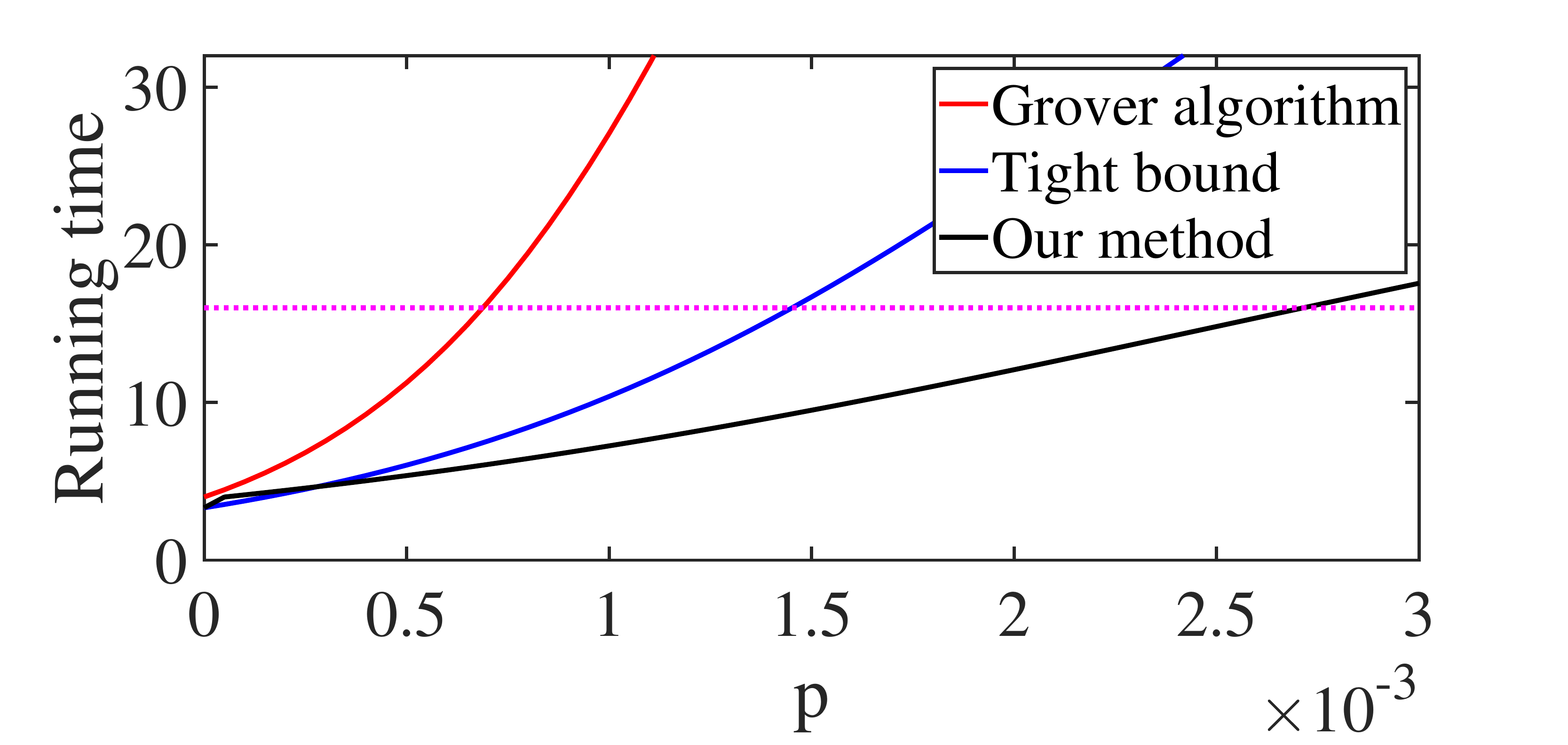}
		\end{minipage}
		\hspace{0.6cm}
		\begin{minipage}{0.245\linewidth}
			\includegraphics[width=1.3\linewidth]{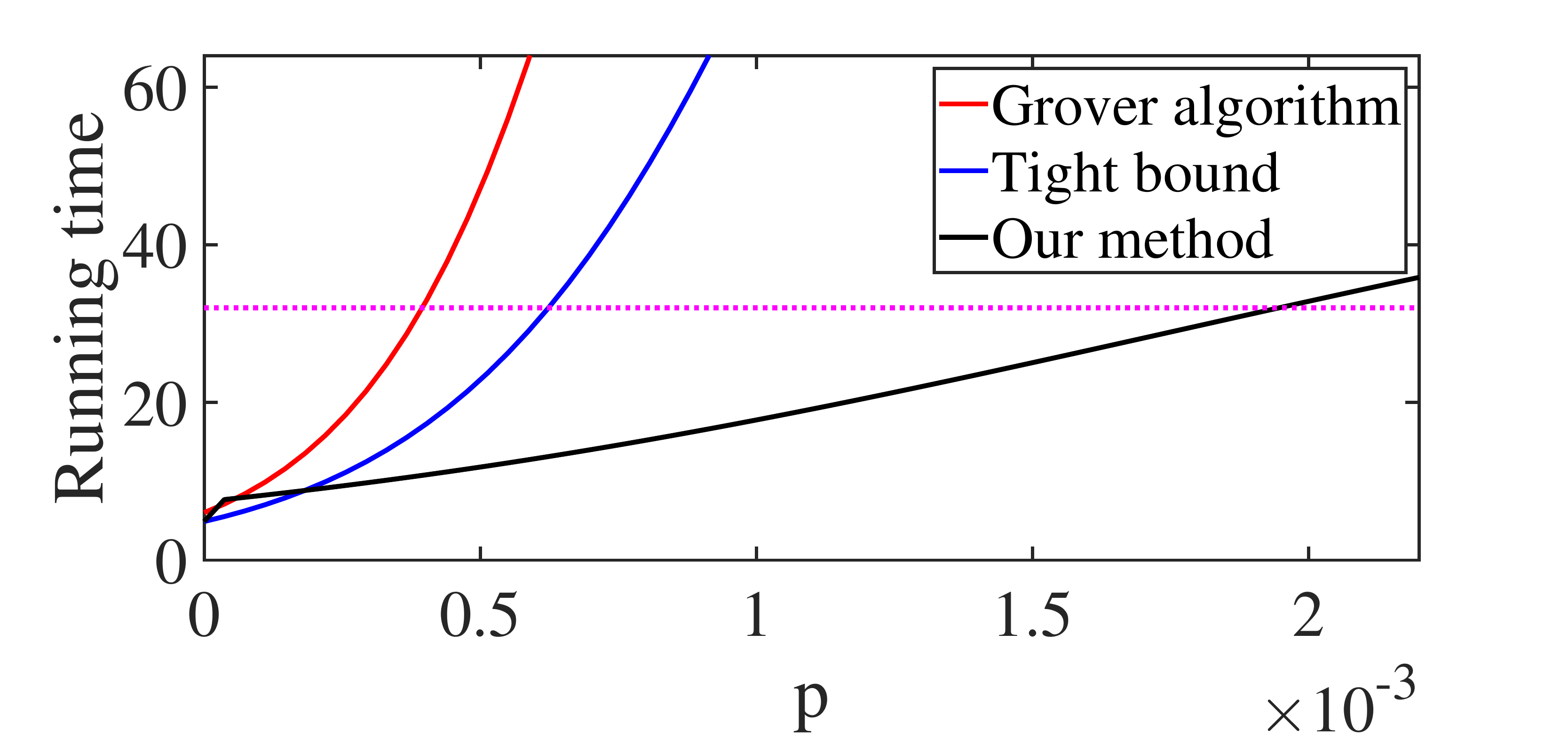}
		\end{minipage}
		\hspace{0.6cm}
		\begin{minipage}{0.245\linewidth}
			\includegraphics[width=1.3\linewidth]{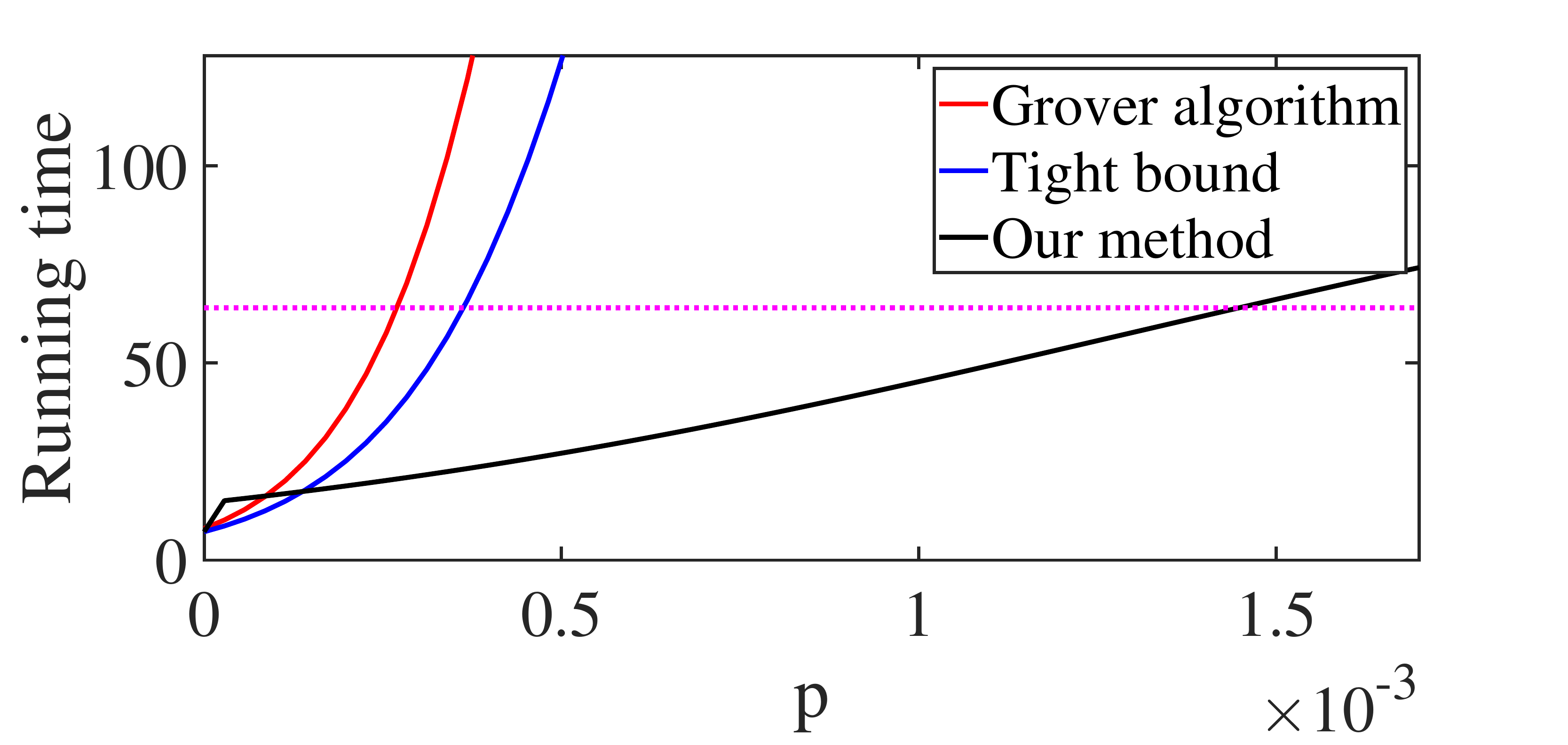}
		\end{minipage}
		\hspace{0.6cm}
		\begin{minipage}{0.245\linewidth}
			\includegraphics[width=1.3\linewidth]{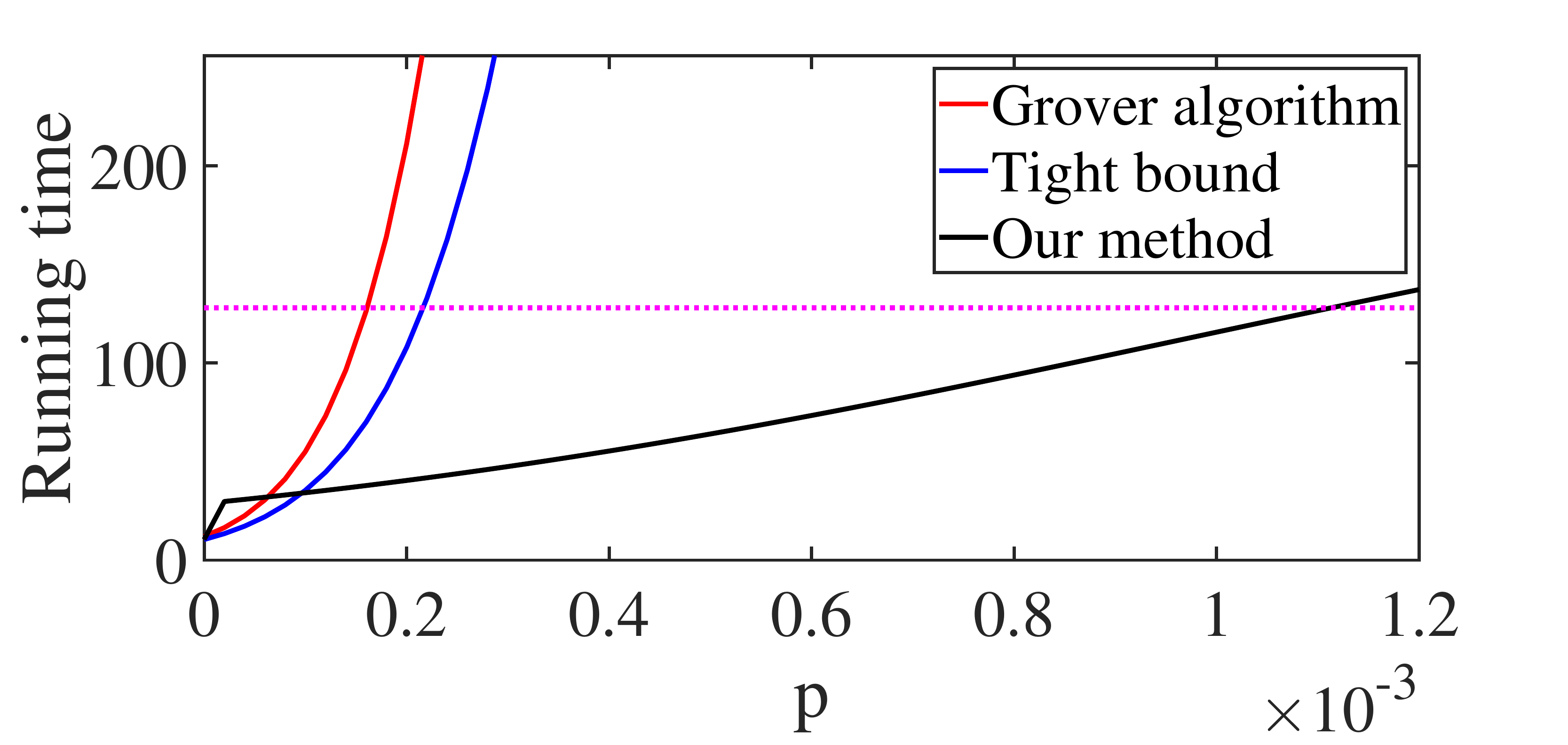}
		\end{minipage}
		\hspace{-1cm}
		
		\leftline{\hspace{-1.7cm}(c) Cross-talk noise.}
		
		\leftline{\hspace{-1.1cm}$n=5$, $t=6$.~~~~~~~~~~~~~~~~~~~~~~~~~~~~~~~$n=6$, $t=7$.~~~~~~~~~~~~~~~~~~~~~~~~~~~~~~~~$n=7$, $t=8$.~~~~~~~~~~~~~~~~~~~~~~~~~~~~~~~$n=8$, $t=9$.}
		
		\hspace{-2.4cm}
		\begin{minipage}{0.245\linewidth}
			\includegraphics[width=1.3\linewidth]{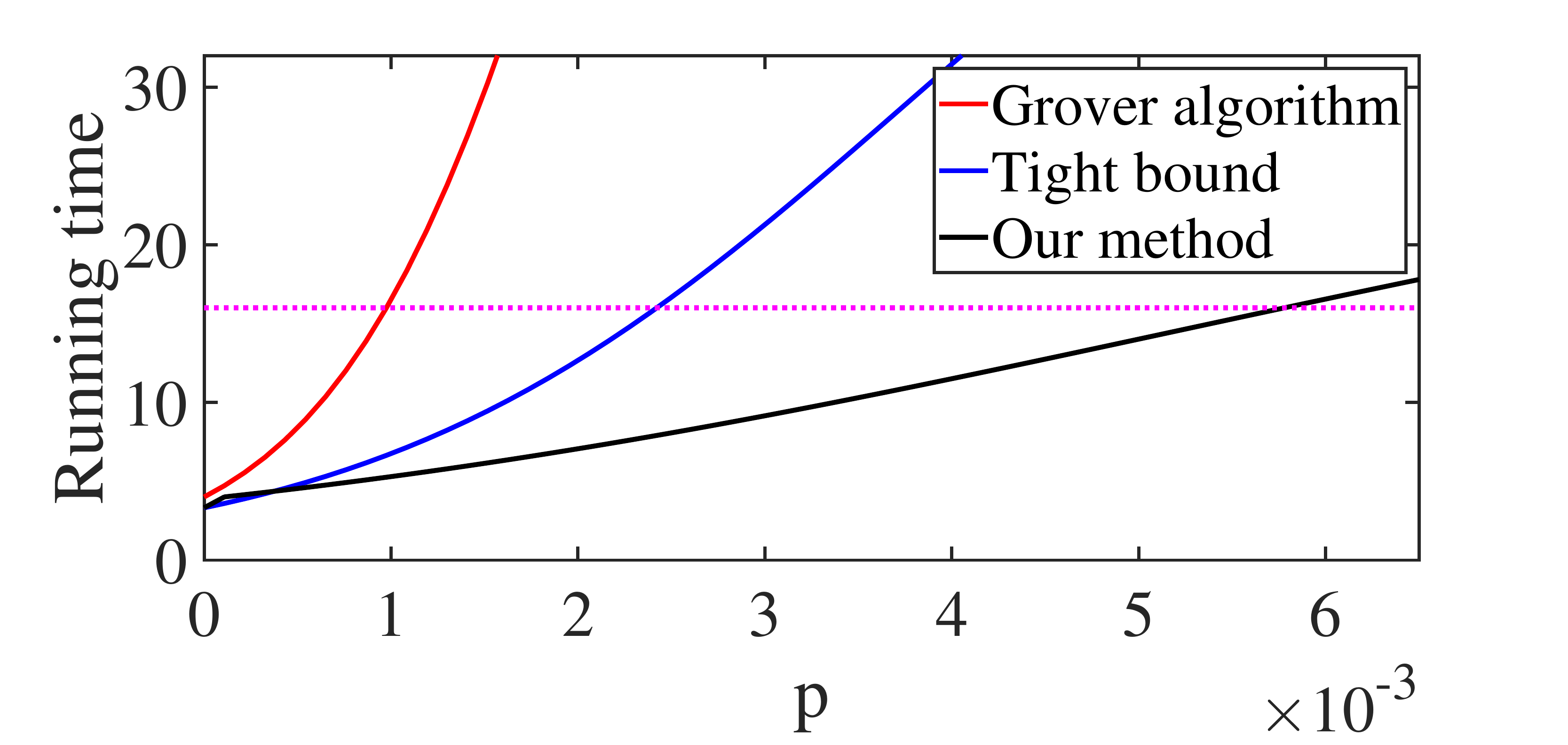}
		\end{minipage}
		\hspace{0.6cm}
		\begin{minipage}{0.245\linewidth}
			\includegraphics[width=1.3\linewidth]{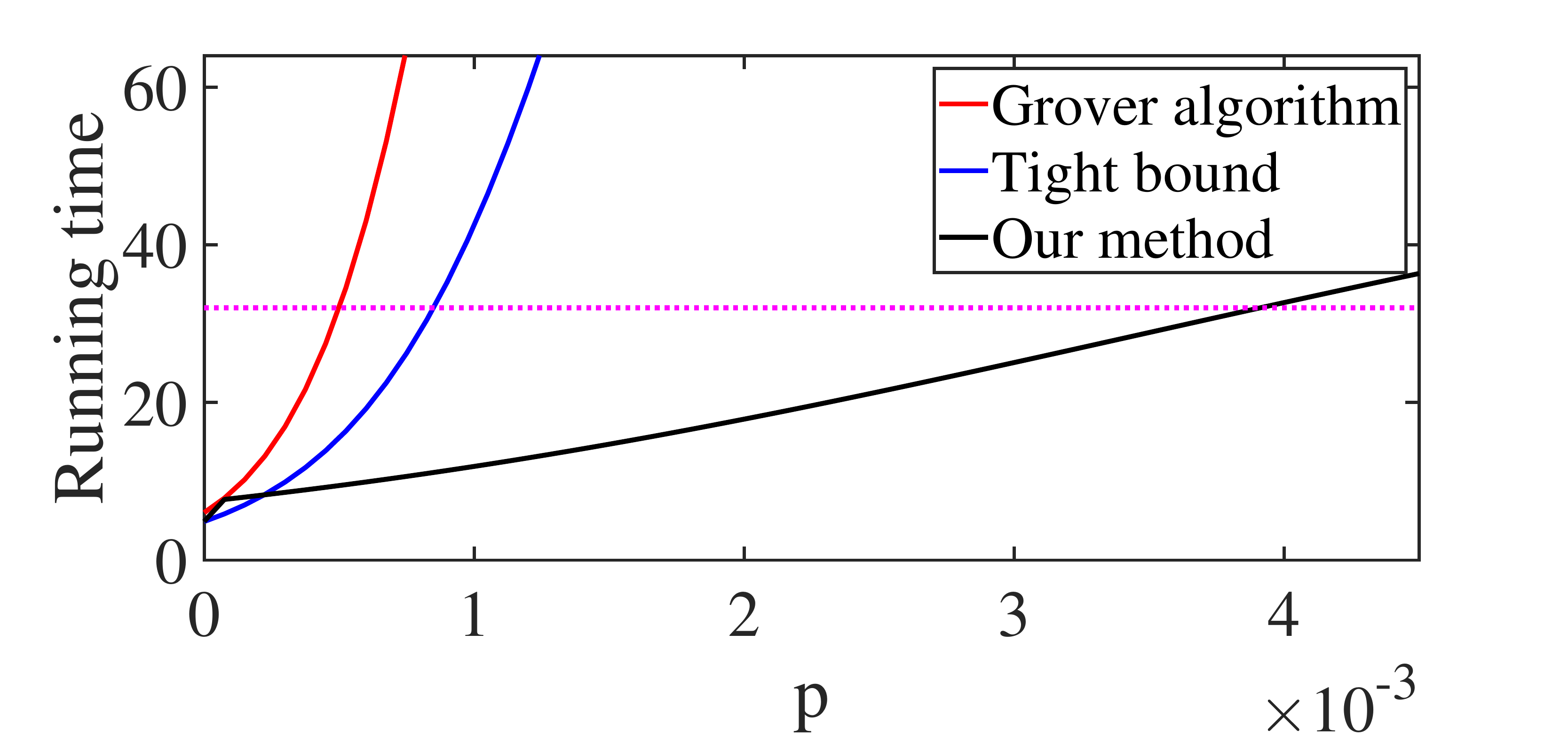}
		\end{minipage}
		\hspace{0.6cm}
		\begin{minipage}{0.245\linewidth}
			\includegraphics[width=1.3\linewidth]{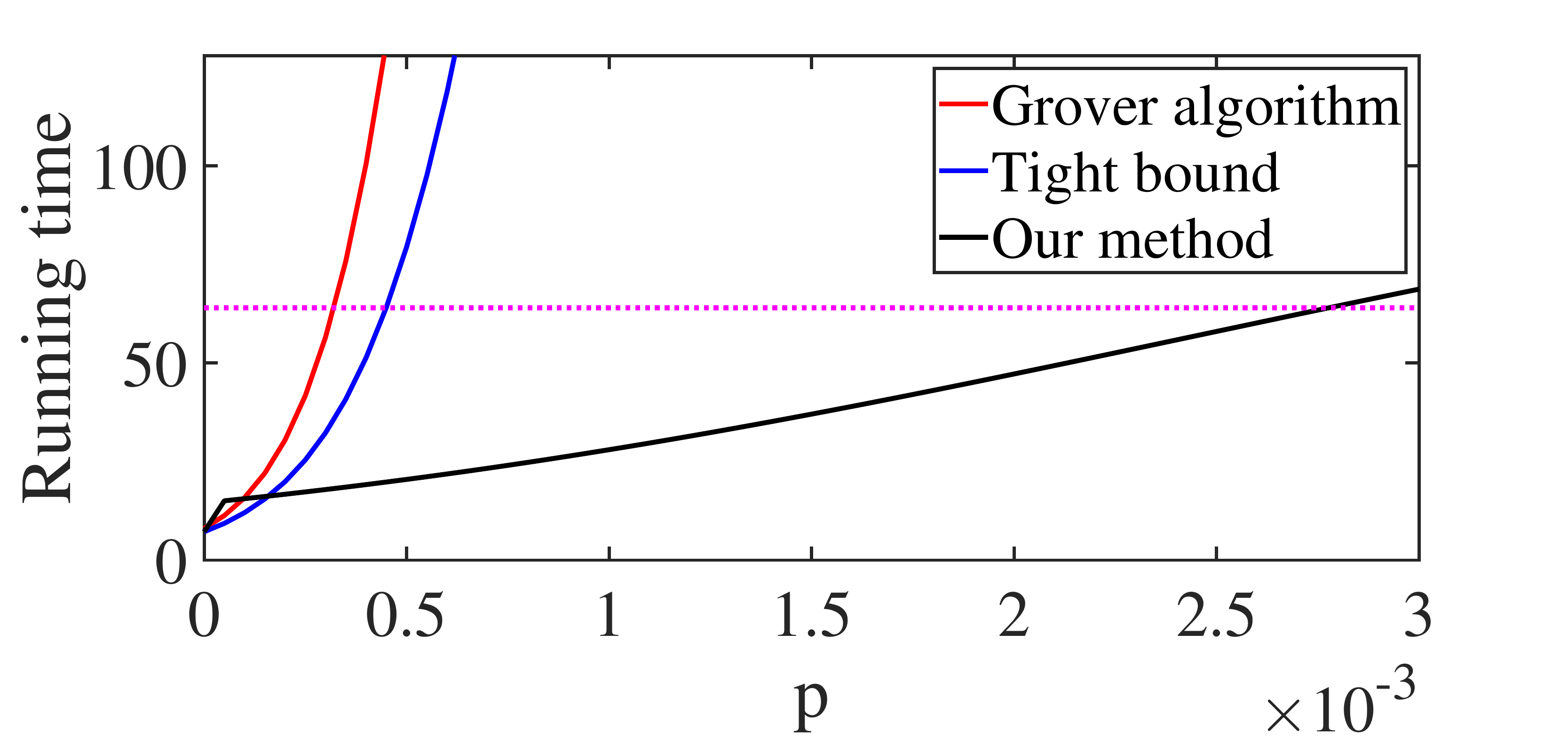}
		\end{minipage}
		\hspace{0.6cm}
		\begin{minipage}{0.245\linewidth}
			\includegraphics[width=1.3\linewidth]{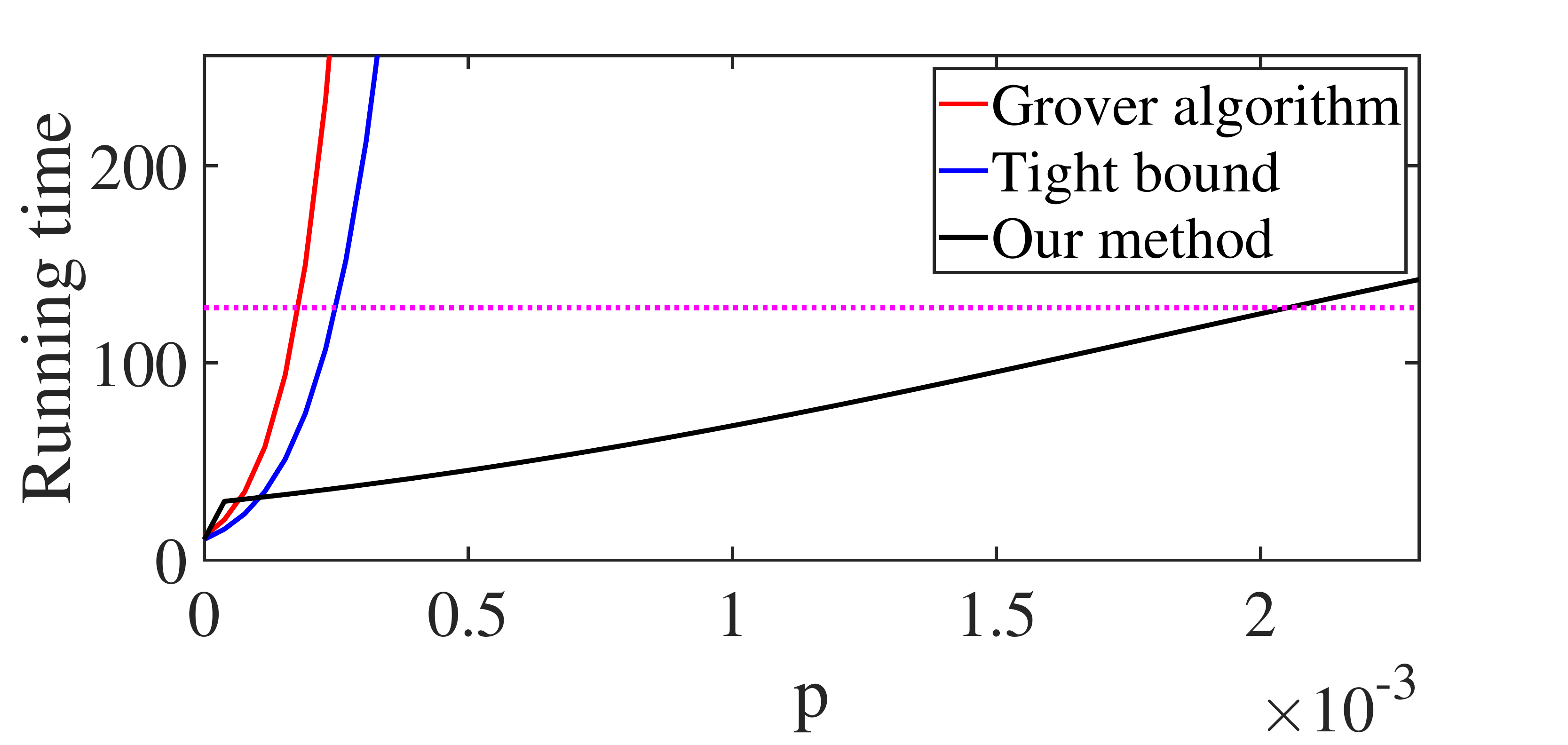}
		\end{minipage}
		\hspace{-1cm}
		
		\leftline{\hspace{-1.7cm}(d) Depolarizing noise.}
		
		\leftline{\hspace{-1.1cm}$n=5$, $t=6$.~~~~~~~~~~~~~~~~~~~~~~~~~~~~~~~$n=6$, $t=7$.}
		
		\hspace{-12cm}
		\begin{minipage}{0.245\linewidth}
			\includegraphics[width=1.3\linewidth]{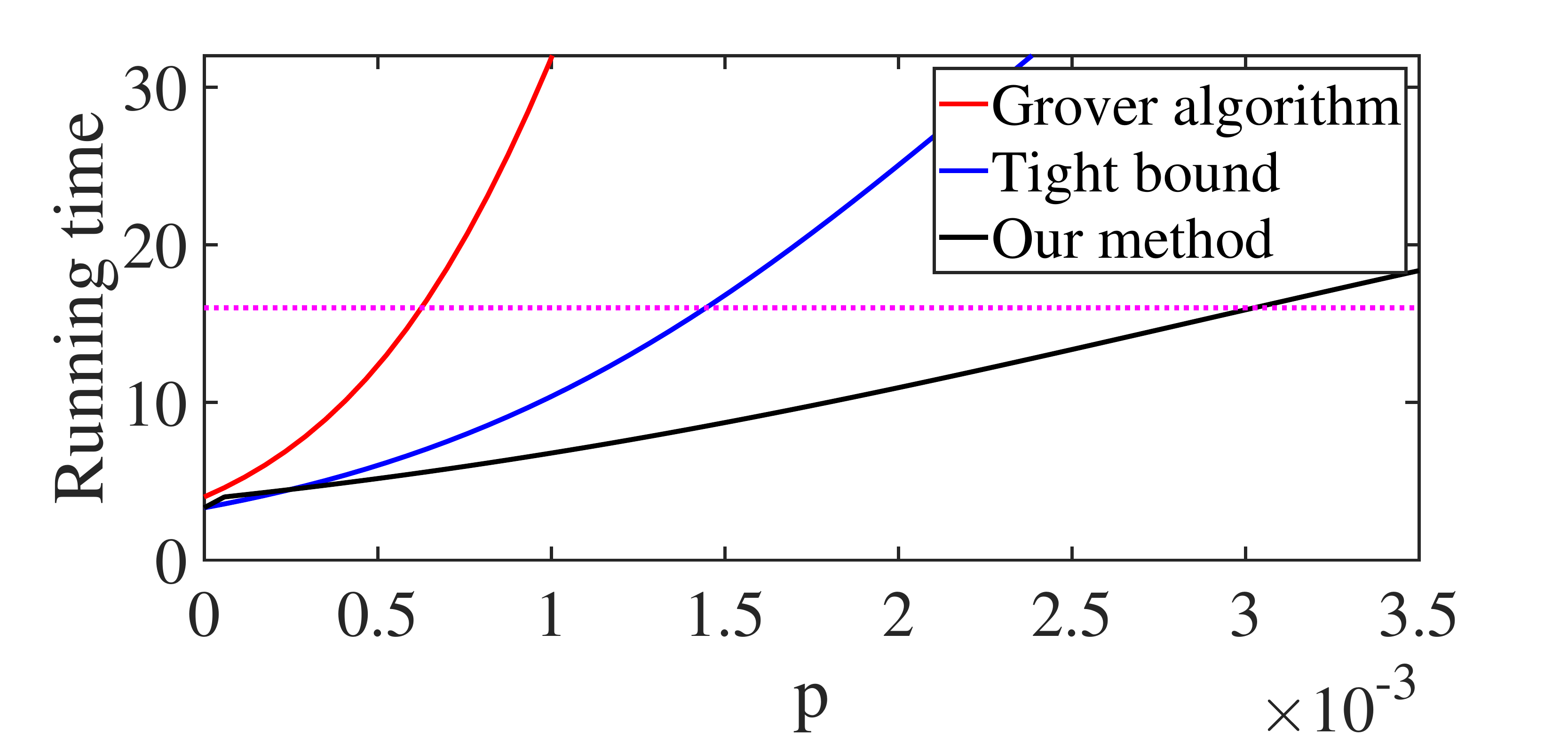}
		\end{minipage}
		\hspace{0.6cm}
		\begin{minipage}{0.245\linewidth}
			\includegraphics[width=1.3\linewidth]{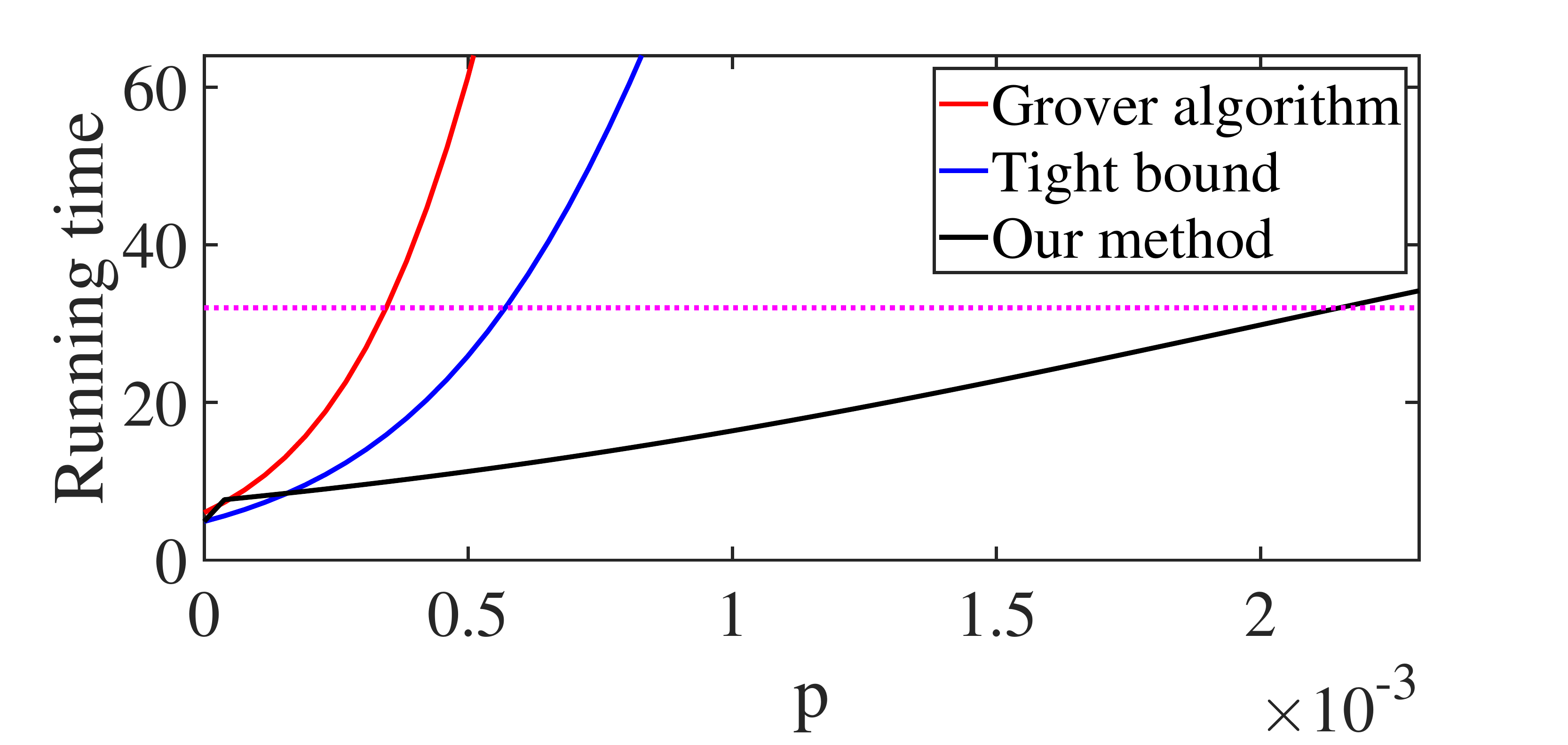}
		\end{minipage}
		
		\leftline{\hspace{-1.7cm}(e) Global depolarizing noise.}
		
		\leftline{\hspace{-1.1cm}$n=5$, $t=6$.~~~~~~~~~~~~~~~~~~~~~~~~~~~~~~~$n=10$, $t=11$.~~~~~~~~~~~~~~~~~~~~~~~~~~~~~$n=15$, $t=16$.~~~~~~~~~~~~~~~~~~~~~~~~~~~~$n=20$, $t=21$.}
		
		\hspace{-2.5cm}
		\begin{minipage}{0.245\linewidth}
			\includegraphics[width=1.3\linewidth]{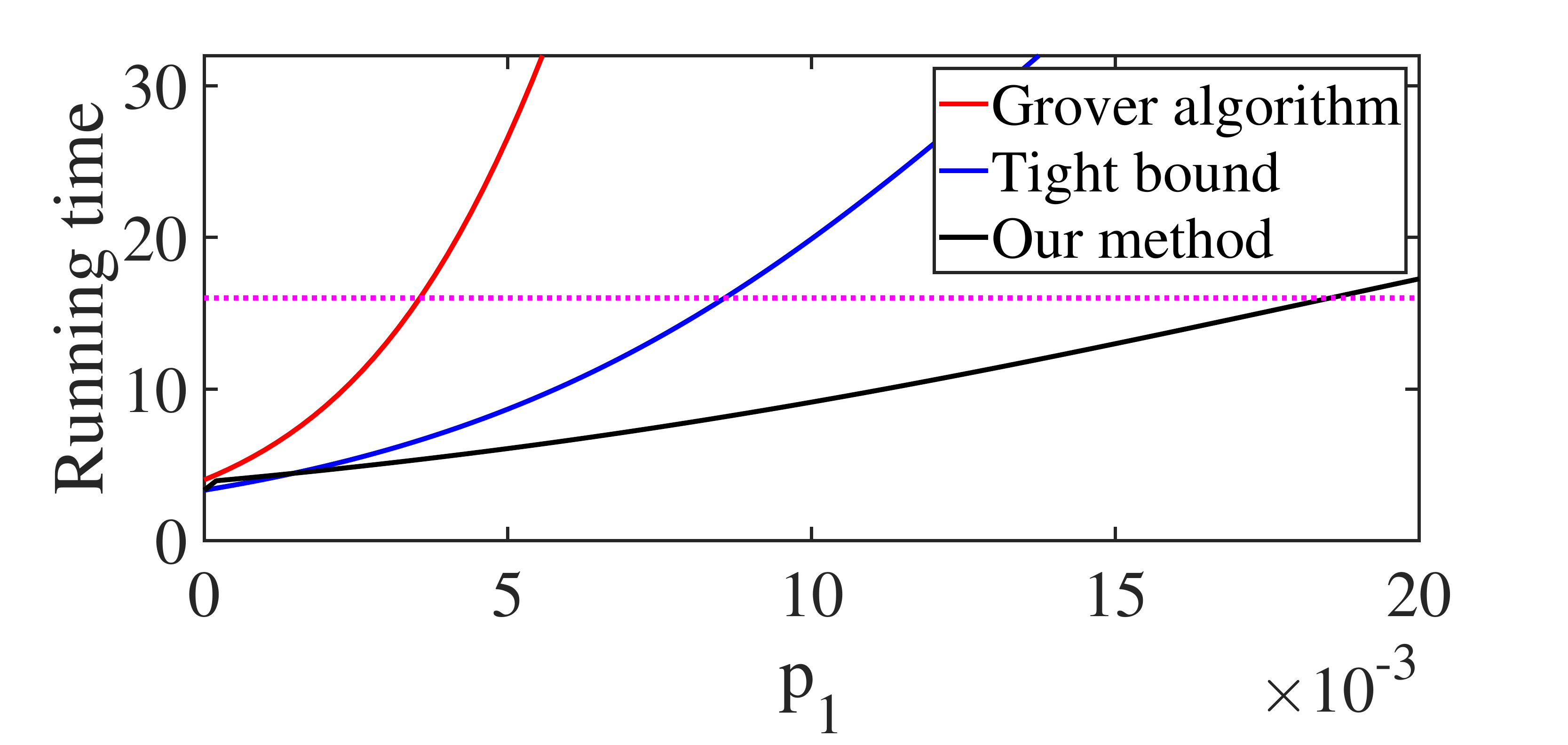}
		\end{minipage}
		\hspace{0.7cm}
		\begin{minipage}{0.245\linewidth}
			\includegraphics[width=1.3\linewidth]{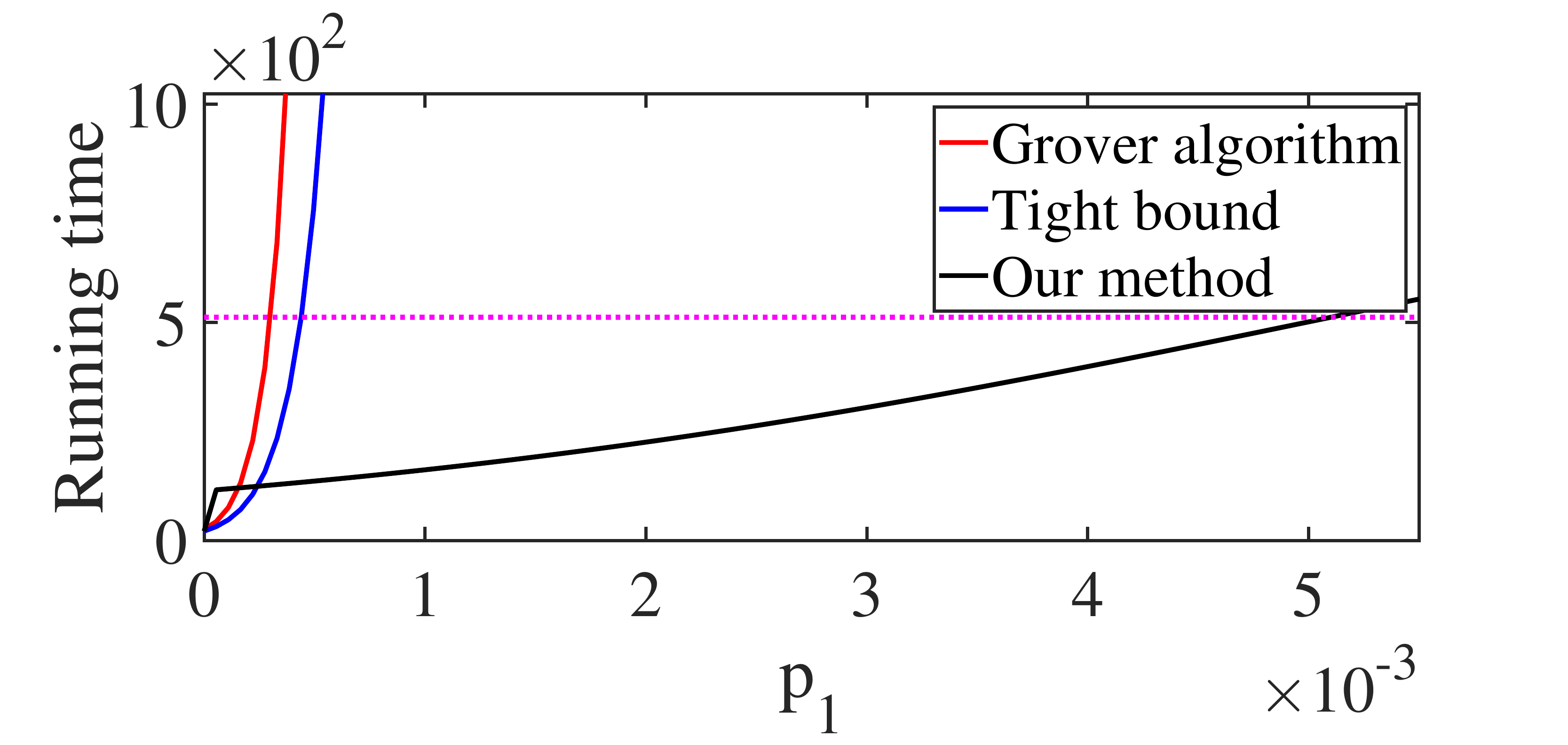}
		\end{minipage}
		\hspace{0.6cm}
		\begin{minipage}{0.245\linewidth}
			\includegraphics[width=1.3\linewidth]{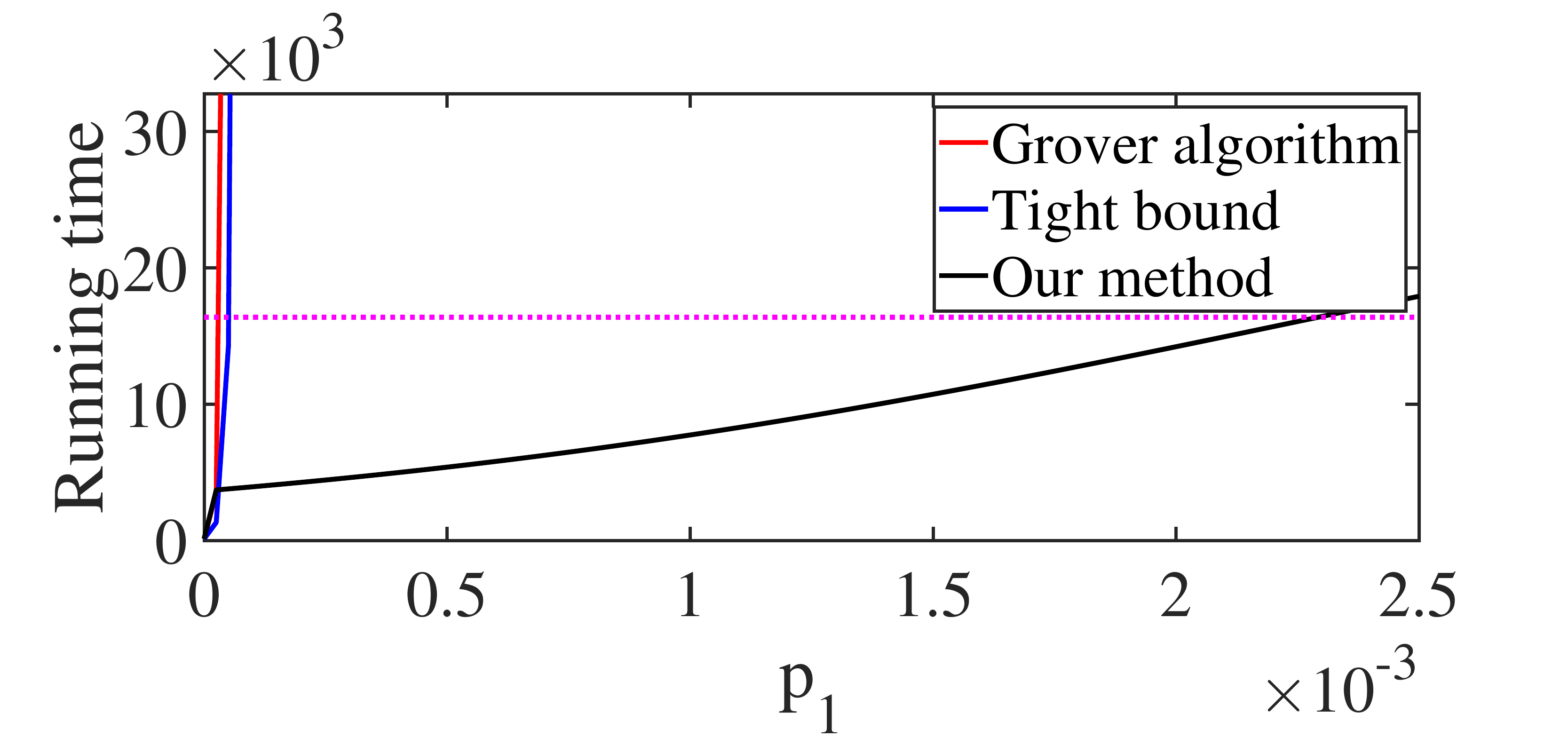}
		\end{minipage}
		\hspace{0.6cm}
		\begin{minipage}{0.245\linewidth}
			\includegraphics[width=1.3\linewidth]{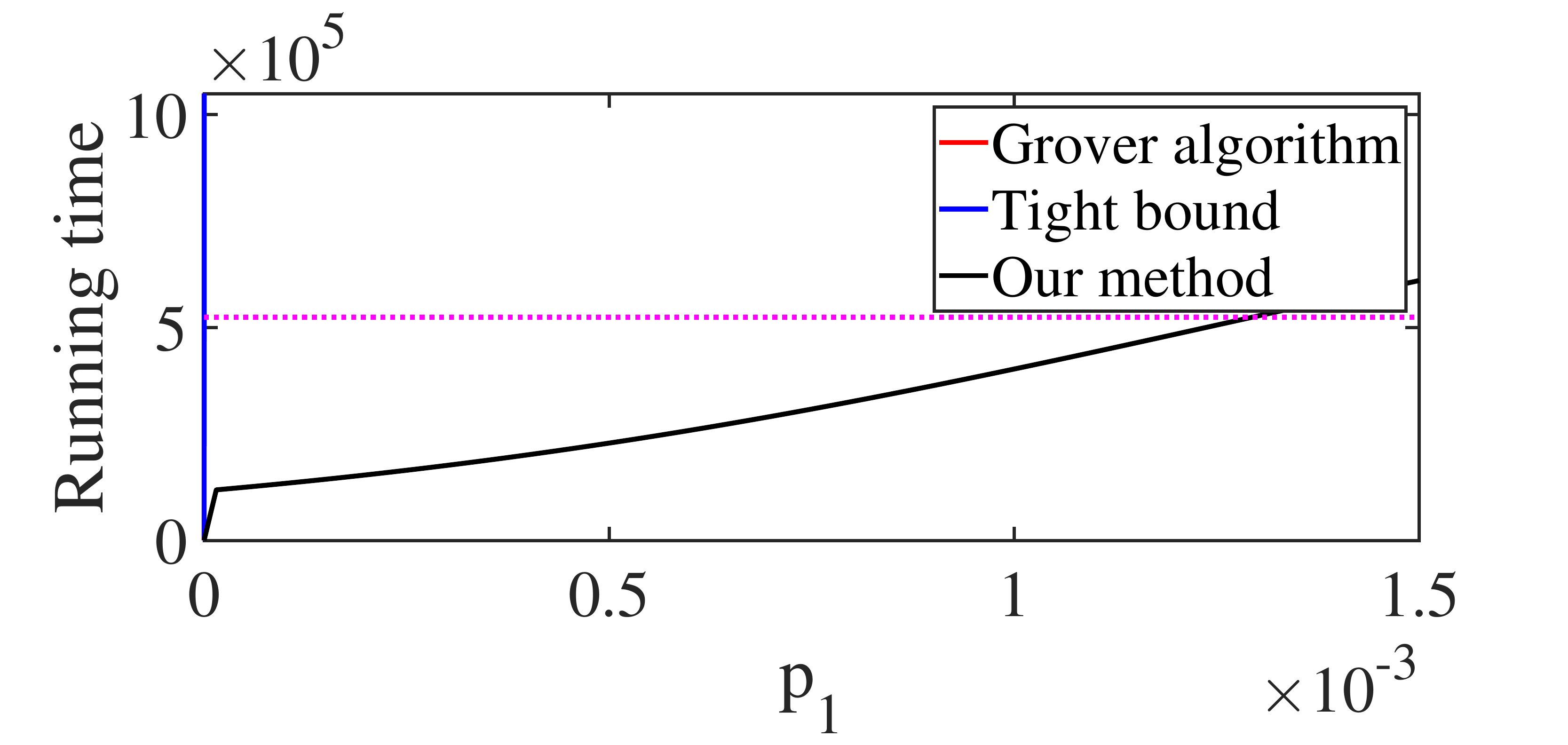}
		\end{minipage}
		\hspace{-1cm}
		
		\caption{\label{running_time}Numerical simulation for running time of Grover's algorithm, tight bound from Ref. \cite{boyer1998tight} and our noise-tolerant method under (a) bit-flip, (b) phase-flip, (c) cross-talk, (d) depolarizing noise and (e) global depolarizing. $n$ is the amount of qubits in register $\alpha$ and $t=n+1$ is the total qubit amount. $p$ is the noise parameter as shown in Appendix B. Pink dashed line is the running time of classically random sampling. The value of $p$ on the intersection between dashed and solid line is the noise threshold $p^\prime$ for quantum advantage.}
	\end{figure*}

	Now we show the details of numerical simulation. We first decompose the Grover circuit with local gates set that contains $\mathcal{O}(t)$ elements. The circuit in Fig. \ref{Grover_circuit} already includes some local gates:
	\begin{align}
		\{\tilde{X_1},~\tilde{X_2},~...,~\widetilde{X_{n+1}};~\tilde{H_1},~\tilde{H_2},~...,~\widetilde{H_{n+1}}\}\label{Grover_local_gates_1}
	\end{align}
	The remaining gates are oracle operator $S_o$ and $(n+1)$-qubit Toffoli gate $\mathrm{Toff}_{\alpha\beta}$ and they are non-local gates. The circuit for $S_o$ changes with different search problems. In this paper we choose the oracle operator as shown in Fig. \ref{oracle_circuit}.
	This oracle can be described in Eq. (\ref{oracle_operator}) that $f(x)=1$ if $x=N-1$ and otherwise $f(x)=0$. In this case, register $\gamma$ does not work and the total amount of qubits is $t=n+1$. Now the only non-local gate in Fig. \ref{Grover_circuit} is $(n+1)$-qubit Toffoli gate and its decomposition is shown in Appendix A.	Combining Eq. (\ref{Grover_local_gates_1}) and (\ref{Grover_local_gates_2}) we obtain the local gates set $\{\tilde{G}^{(i)}_a\}$. We continually apply local gates on qubits according to the decomposition above. Noise model is shown in Appendix B. We numerically simulate the Grover circuit to obtain the running times of Grover's algorithm, tight bound from Ref. \cite{boyer1998tight} and our noise-tolerant method under bit-flip, phase-flip, cross-talk, depolarizing and global depolarizing noise as shown in Fig. \ref{running_time}.
	\begin{figure*}
		\leftline{\hspace{-1.5cm}(a) Bit-flip noise.~~~~~~~~~~~~~~~~~~~~~~~~~~~~~~~~~~~~~~~~~~~~~~~~~~~~~~~~~~~~~~~~~~~~~~~~~(b) Phase-flip noise.}
		
		\hspace{-2.4cm}
		\begin{minipage}{0.245\linewidth}
			\includegraphics[width=1.3\linewidth]{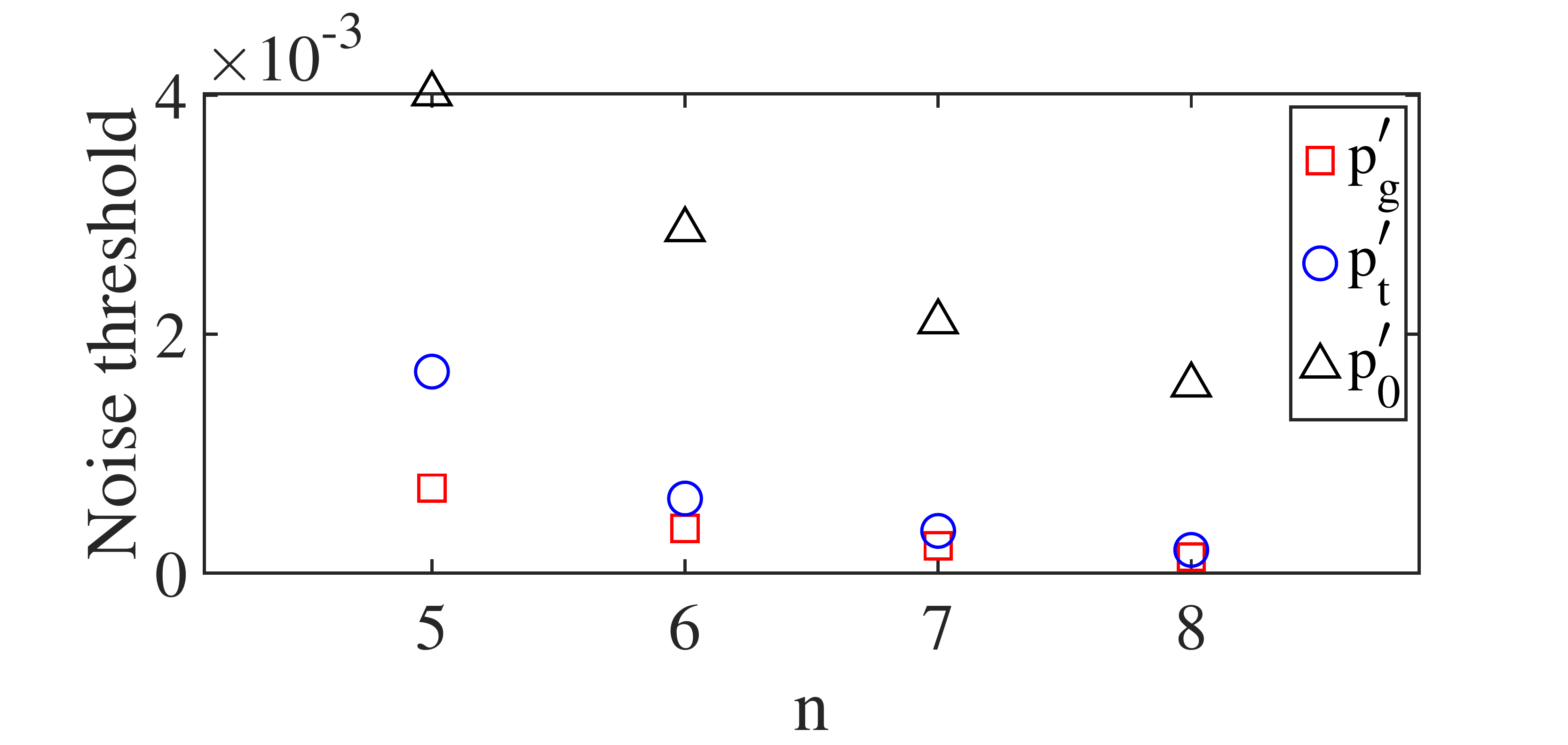}
		\end{minipage}
		\hspace{0.6cm}
		\begin{minipage}{0.245\linewidth}
			\includegraphics[width=1.3\linewidth]{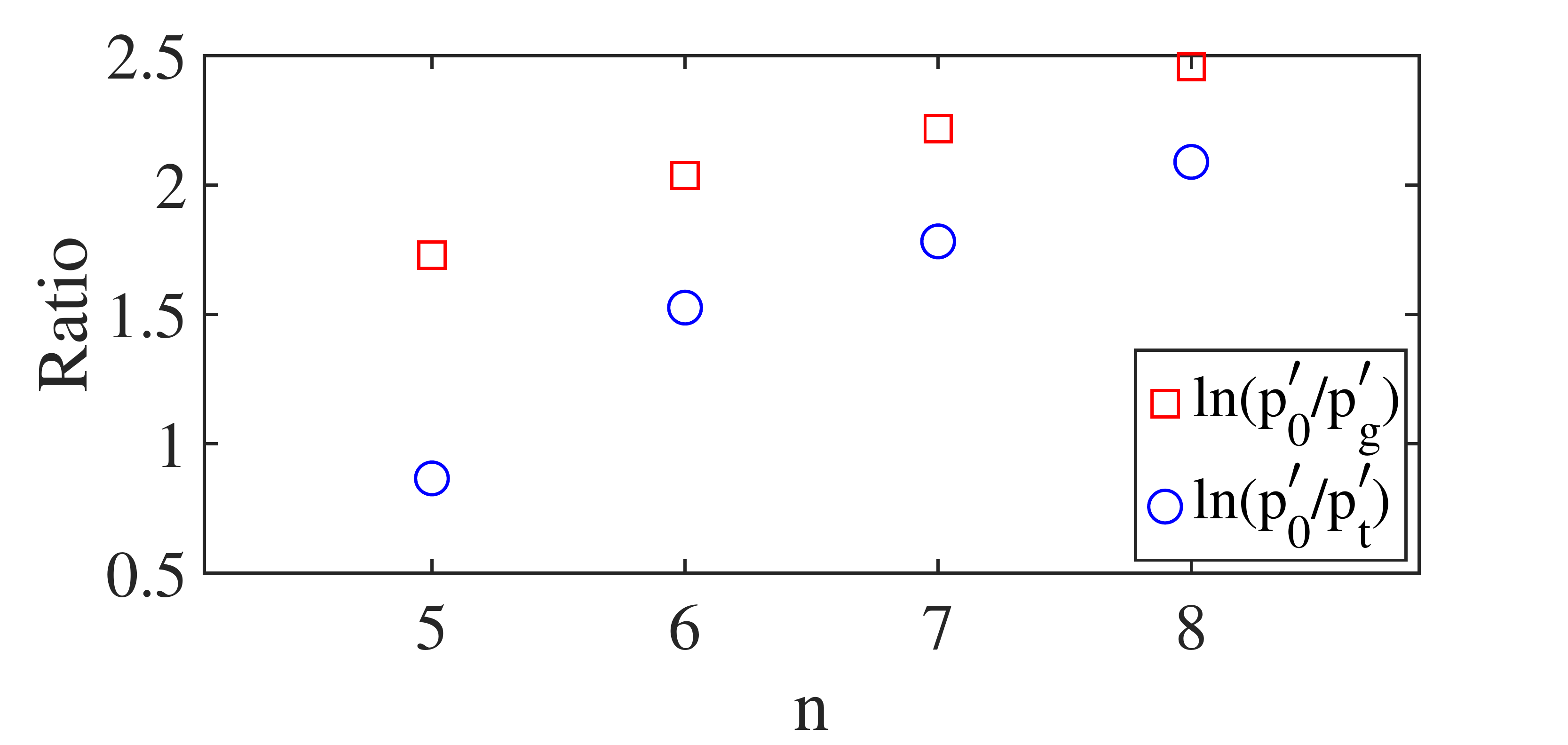}
		\end{minipage}
		\hspace{0.6cm}
		\begin{minipage}{0.245\linewidth}
			\includegraphics[width=1.3\linewidth]{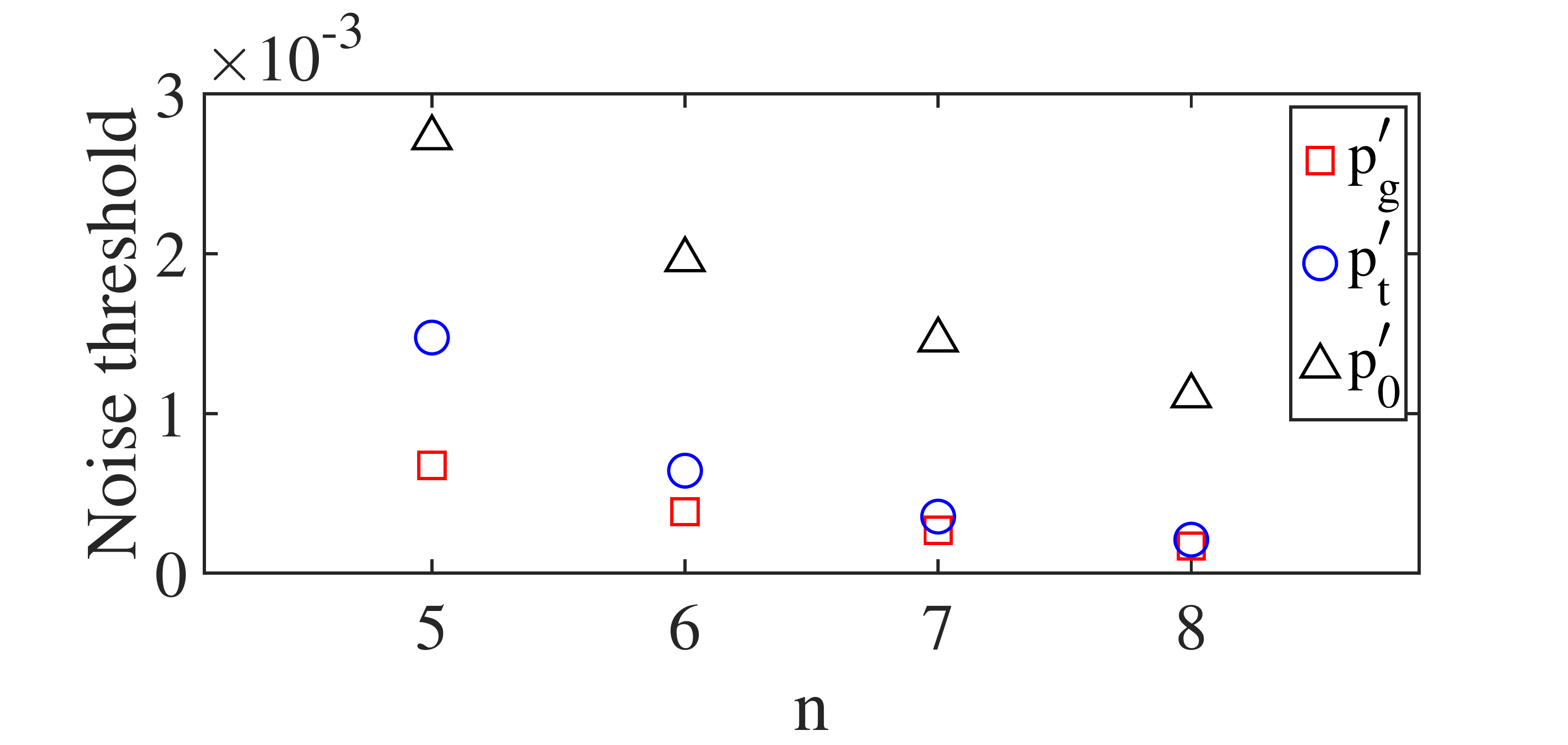}
		\end{minipage}
		\hspace{0.6cm}
		\begin{minipage}{0.245\linewidth}
			\includegraphics[width=1.3\linewidth]{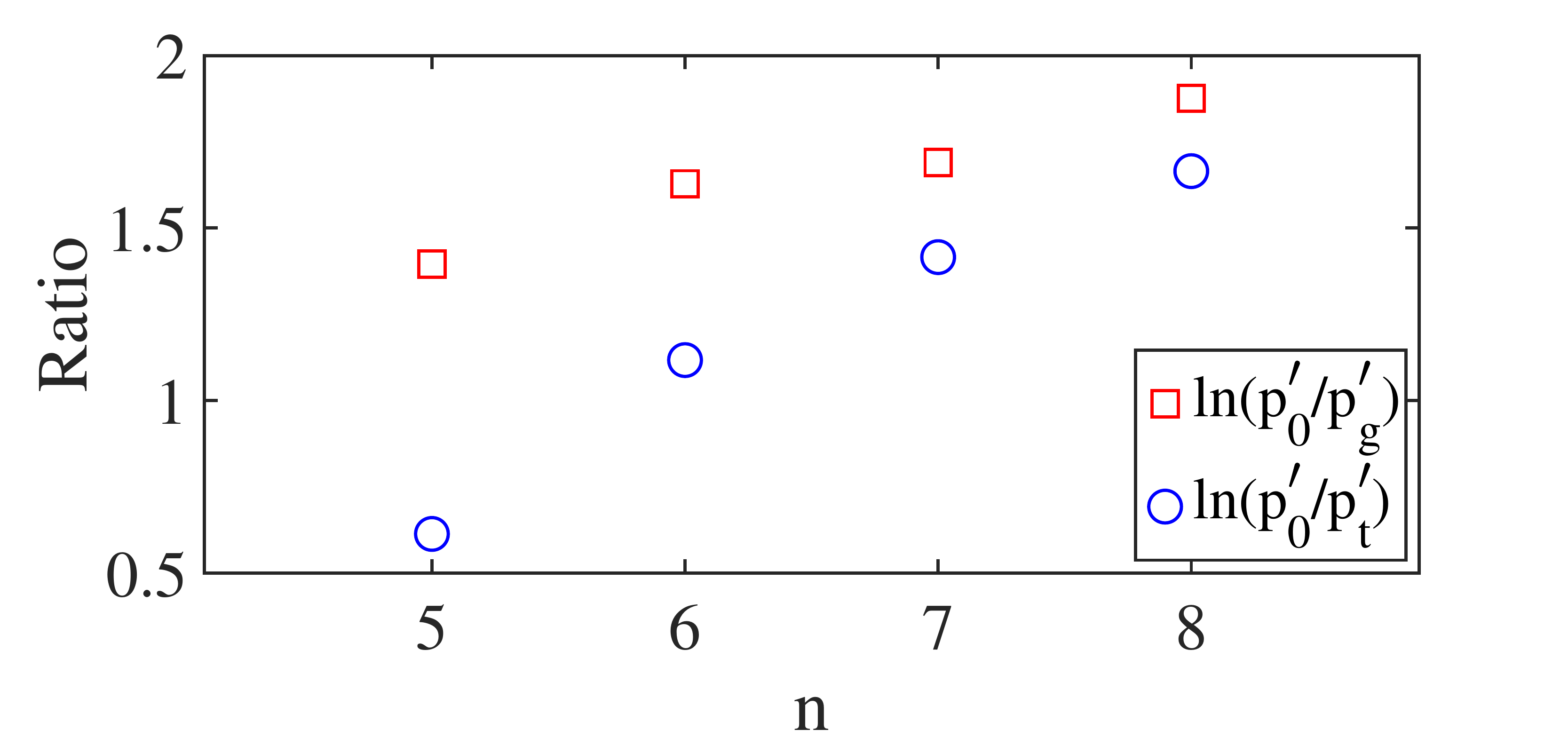}
		\end{minipage}
		\hspace{-1cm}
		
		\leftline{\hspace{-1.5cm}(c) Cross-talk noise.~~~~~~~~~~~~~~~~~~~~~~~~~~~~~~~~~~~~~~~~~~~~~~~~~~~~~~~~~~~~~~~~~~~~~~(d) Depolarizing noise.}
		
		\hspace{-2.4cm}
		\begin{minipage}{0.245\linewidth}
			\includegraphics[width=1.3\linewidth]{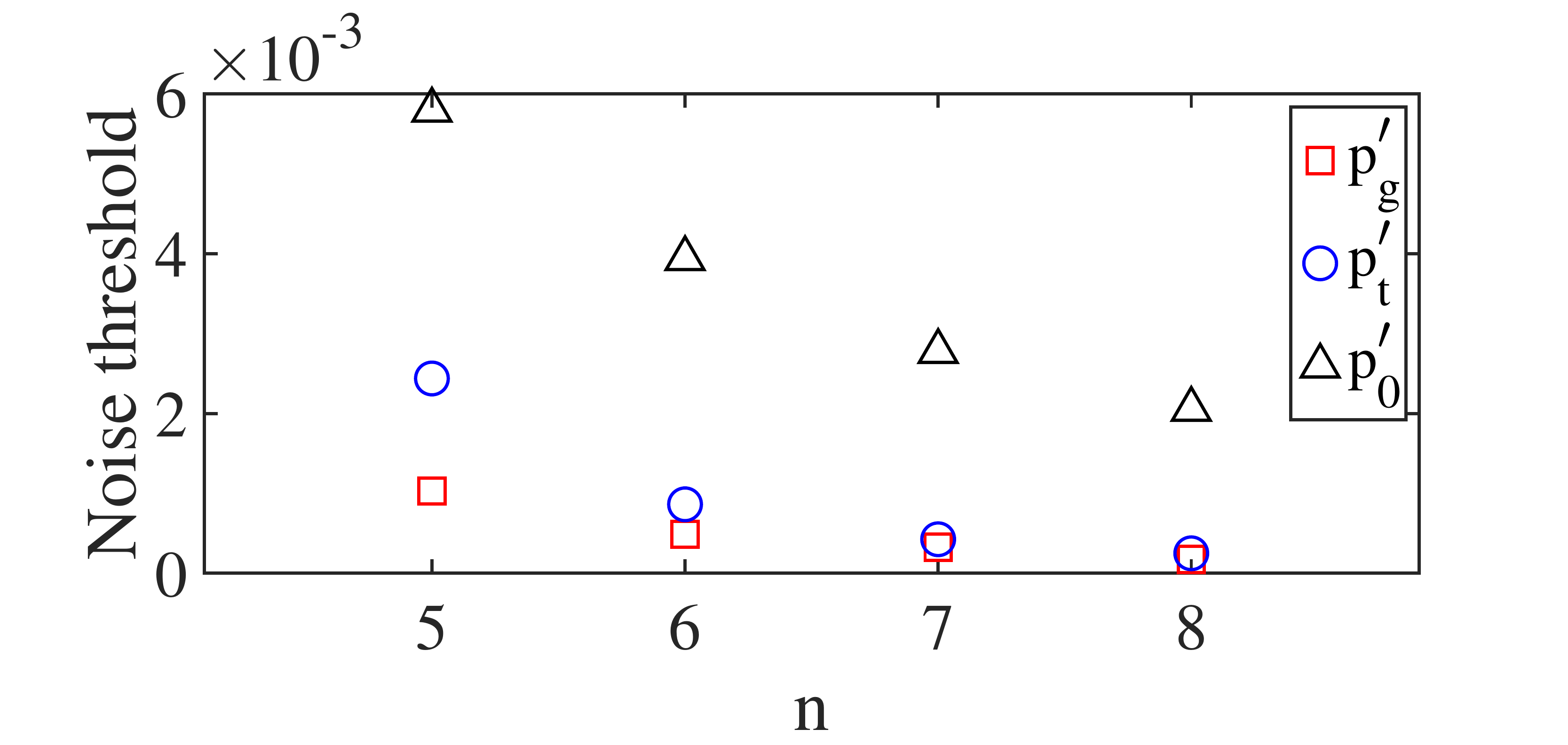}
		\end{minipage}
		\hspace{0.6cm}
		\begin{minipage}{0.245\linewidth}
			\includegraphics[width=1.3\linewidth]{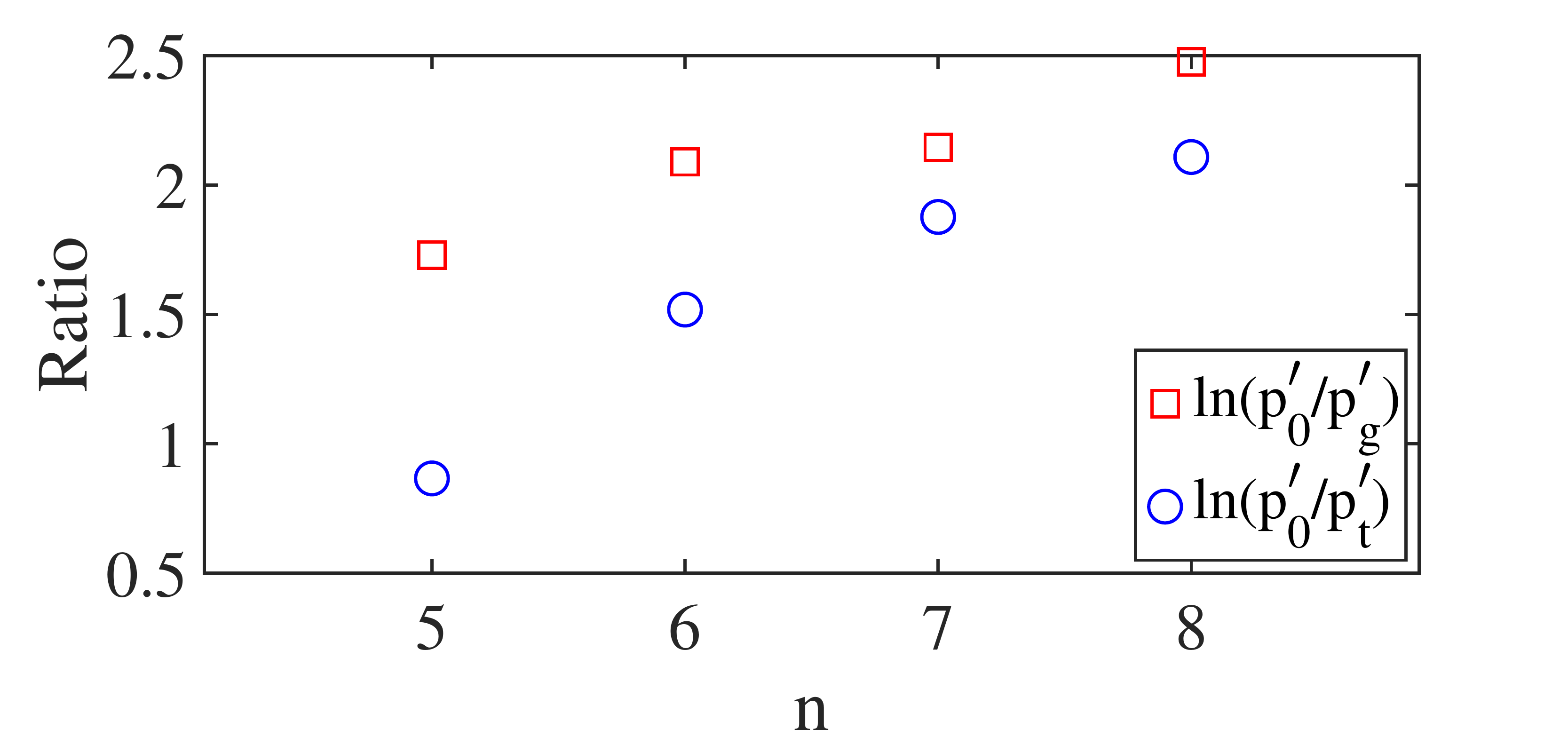}
		\end{minipage}
		\hspace{0.6cm}
		\begin{minipage}{0.245\linewidth}
			\includegraphics[width=1.3\linewidth]{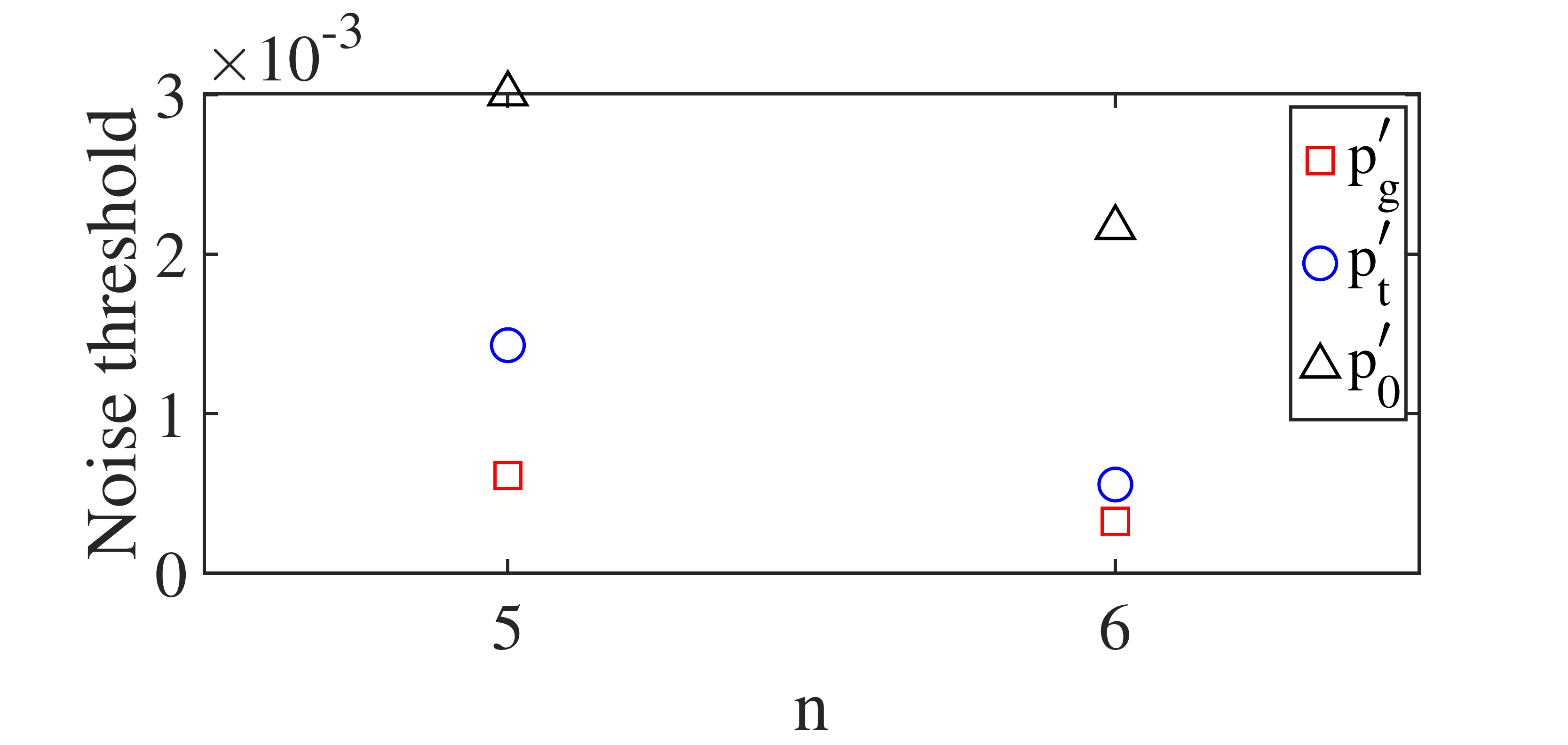}
		\end{minipage}
		\hspace{0.6cm}
		\begin{minipage}{0.245\linewidth}
			\includegraphics[width=1.3\linewidth]{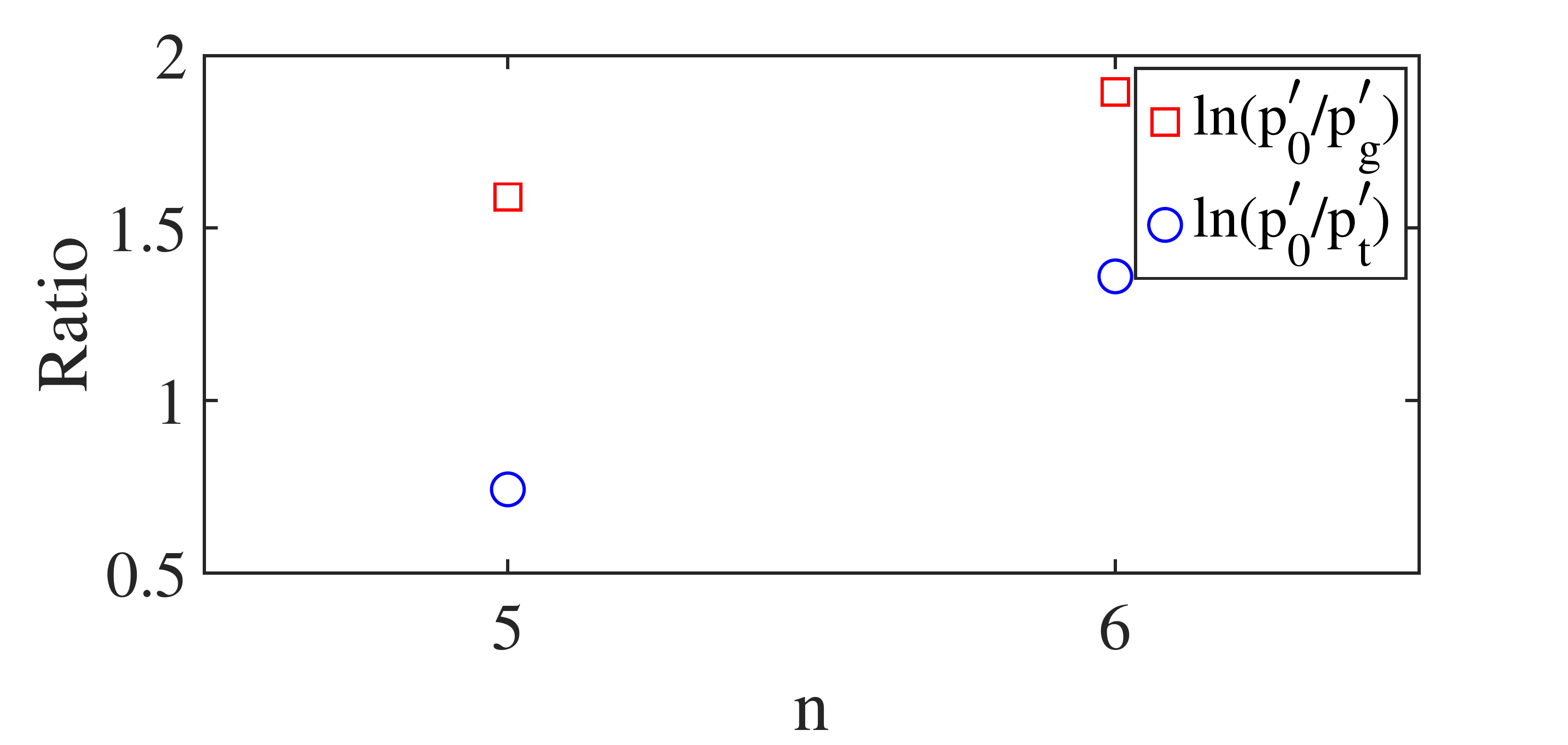}
		\end{minipage}
		\hspace{-1cm}
		
		\leftline{\hspace{-1.5cm}(e) Global depolarizing noise.}
		
		\hspace{-6.6cm}
		\begin{minipage}{0.245\linewidth}
			\includegraphics[width=1.3\linewidth]{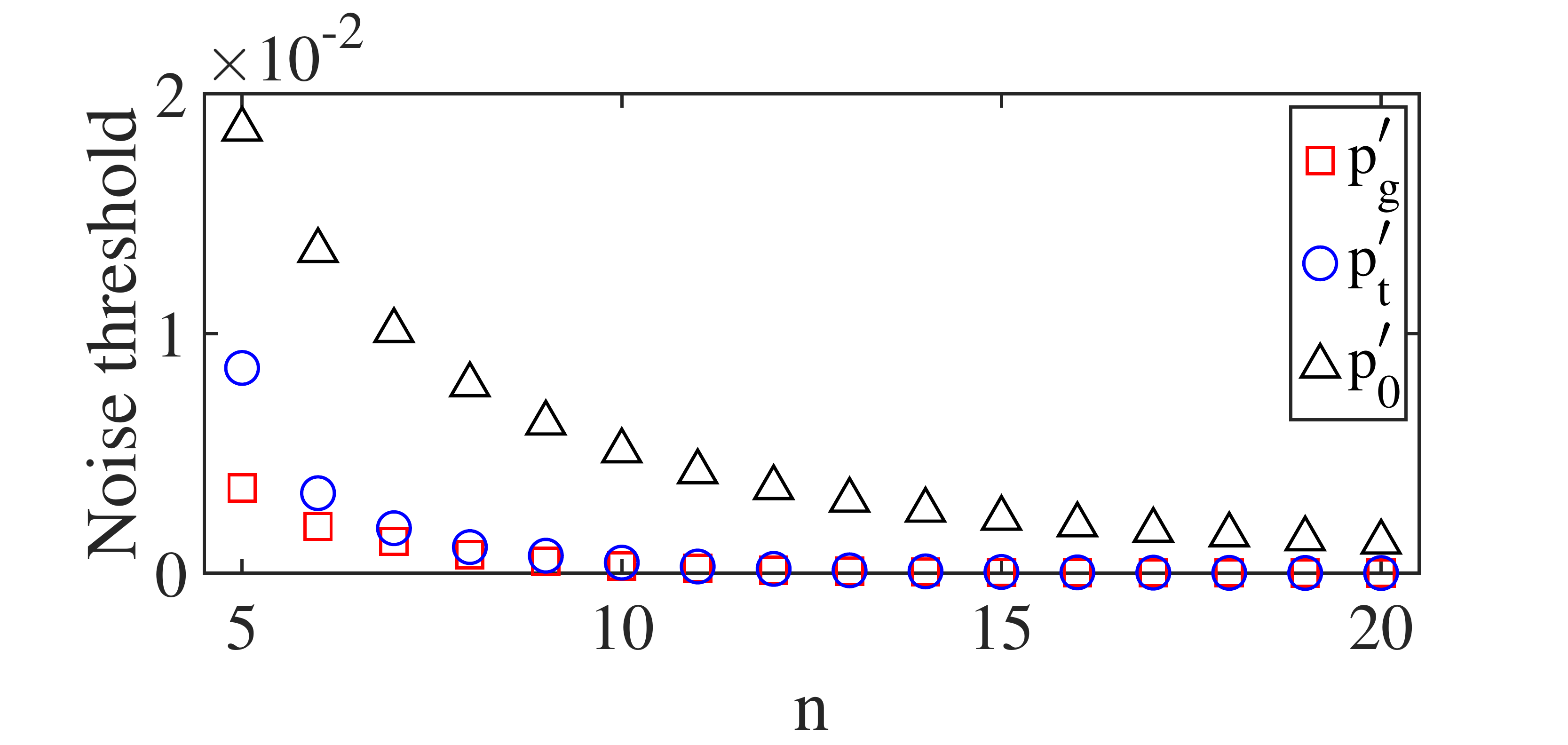}
		\end{minipage}
		\hspace{0.6cm}
		\begin{minipage}{0.245\linewidth}
			\includegraphics[width=1.3\linewidth]{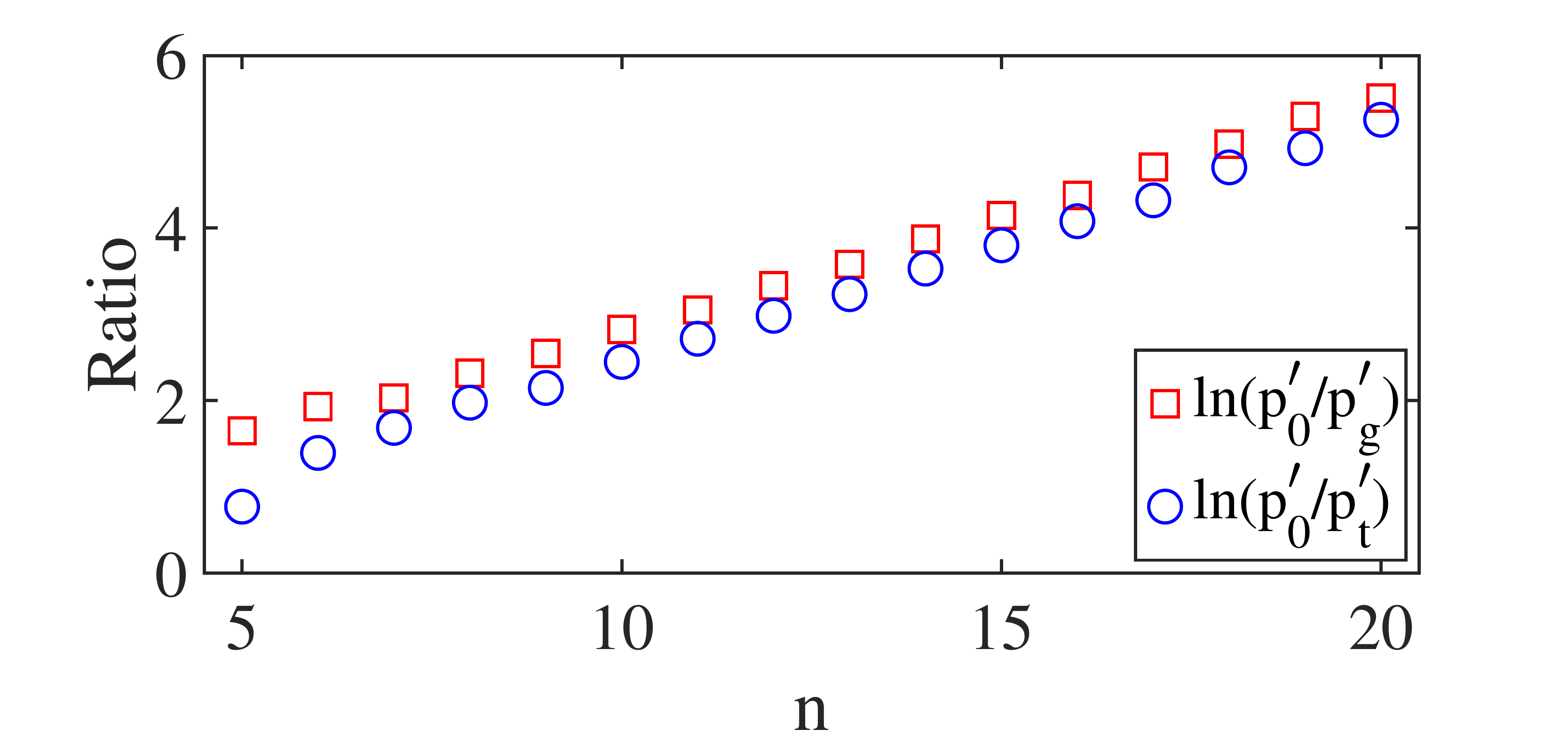}
		\end{minipage}
		\hspace{0.7cm}
		\begin{minipage}{0.245\linewidth}
			\includegraphics[width=1.3\linewidth]{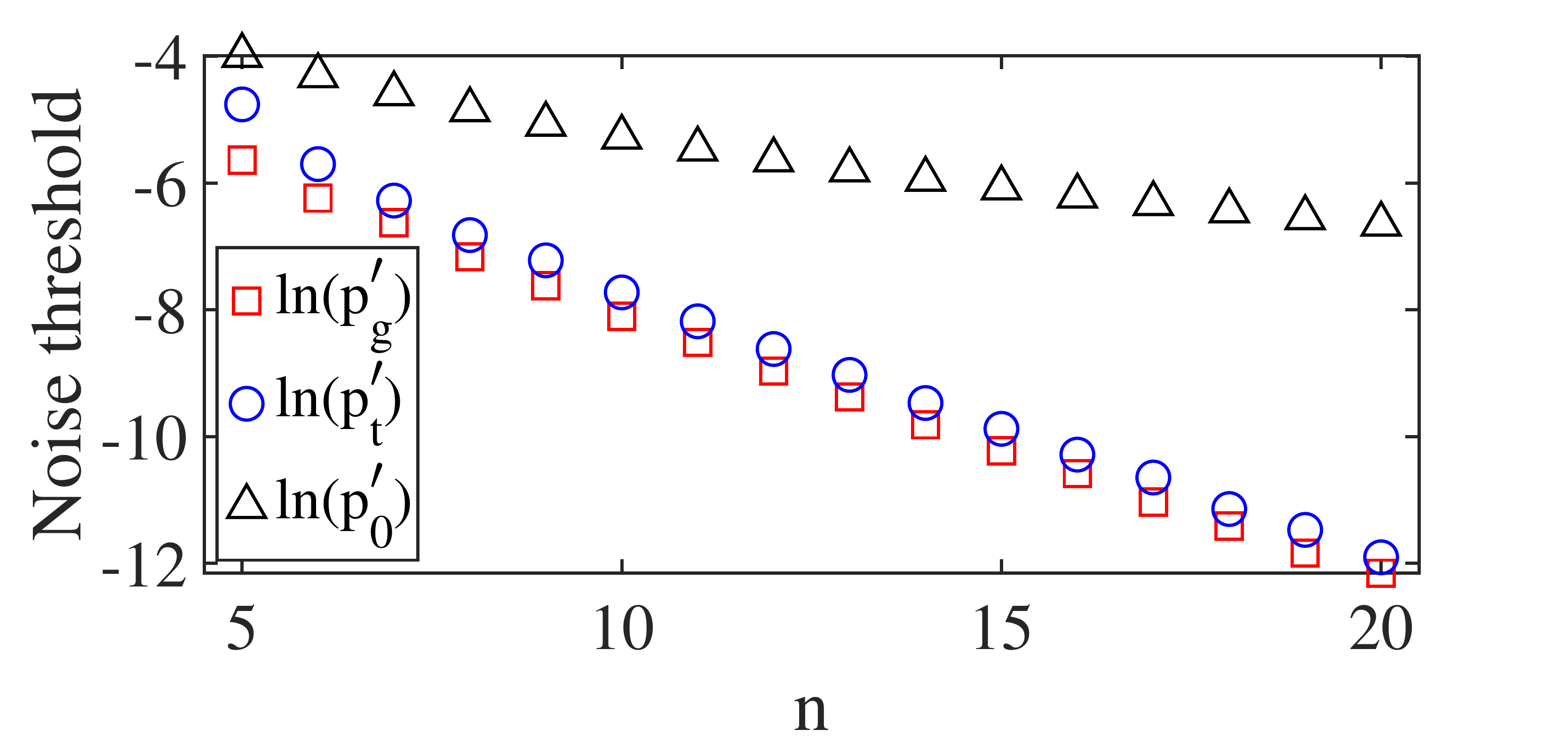}
		\end{minipage}
		
		\caption{\label{noise_threshold}(a) The noise threshold $p^\prime_g$, $p^\prime_t$ and $p^\prime_0$ of Grover's algorithm, tight bound from Ref. \cite{boyer1998tight} and our noise-tolerant method under (a) bit-flip, (b) phase-flip, (c) cross-talk, (d) depolarizing and (e) global depolarizing noise. Our noise-tolerant method exponentially improves the noise threshold of quantum advantage. We add a third figure for global depolarizing noise in sugfigure (e). It shows that $p^\prime_g$ and $p^\prime_t$ exponentially decay while $p^\prime_0$ does not.}
	\end{figure*}
	When the running time of Grover circuit equals to the running time of classical random sampling, the value of $p$ at that point is the noise threshold $p^\prime$ for quantum advantage. The noise thresholds of Grover's algorithm, tight bound from Ref. \cite{boyer1998tight} and our noise-tolerant method are shown in Fig. \ref{noise_threshold}. In conclusion, Fig. \ref{running_time} shows that our noise-tolerant method significantly reduces the running time of Grover's algorithm. Fig. \ref{noise_threshold} shows that the noise threshold for quantum advantage is improved by an exponential factor with qubit amount rise.
	
	\section{Conclusion}
	We present a noise-tolerant method that exponentially improves the noise threshold for quantum advantage of Grover's algorithm. This is a significant progress for realizing quantum advantage in physical systems.
	
	\begin{acknowledgments}
	We acknowledge the financial support in part by National Natural Science Foundation of China grant No.11974204 and No.12174215.
	\end{acknowledgments}

	\appendix
	
	\section{Decompose the $(n+1)$-qubit Toffoli gate}
	
\begin{figure}[htbp]
		\begin{minipage}{1.05\linewidth}
			\vspace{0.3cm}
			\includegraphics[width=1\linewidth]{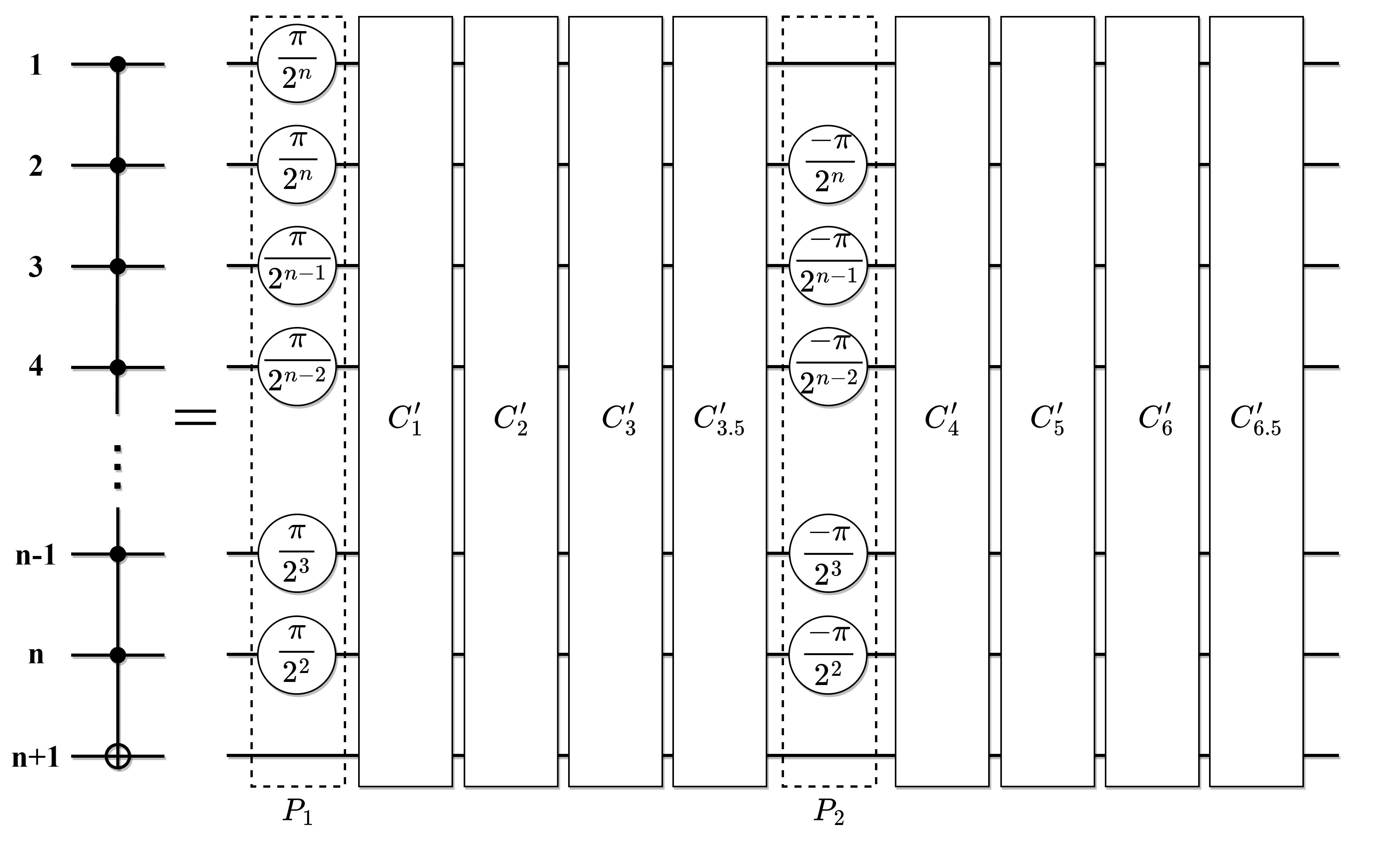}
		\end{minipage}
		\caption{\label{Toffoli_circuit}The $(n+1)$-qubit Toffoli gate is divided into ten parts. $P_1$ and $P_2$ only contain $\theta$-phase gates $S(\theta)$ that are already local gates. The circuit depth of $P_1$ or $P_2$ is $d_{P_1}=d_{P_2}=1$. The decomposition for other parts are shown in Fig. \ref{parts_local_gates}.}
	\end{figure}

\begin{figure*}[htbp]
		\leftline{~$C_1^\prime:~~d_{C_1^\prime}=4n-6$\hspace{6.15cm}$C_4^\prime:~~d_{C_4^\prime}=4n-10$}
		\begin{minipage}{0.49\linewidth}
			\vspace{0.2cm}
			\includegraphics[width=1\linewidth]{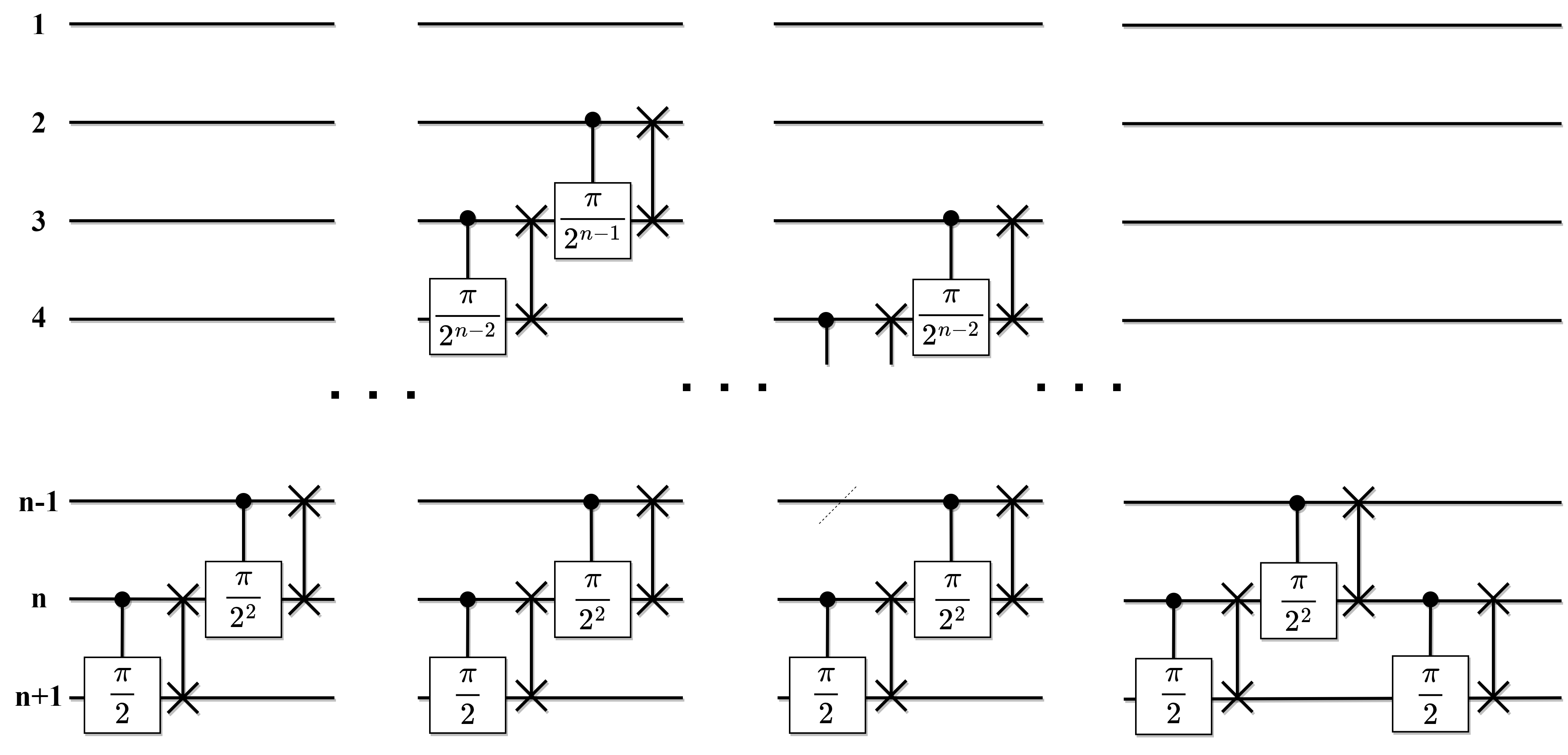}
		\end{minipage}
		\begin{minipage}{0.49\linewidth}
			\includegraphics[width=1\linewidth]{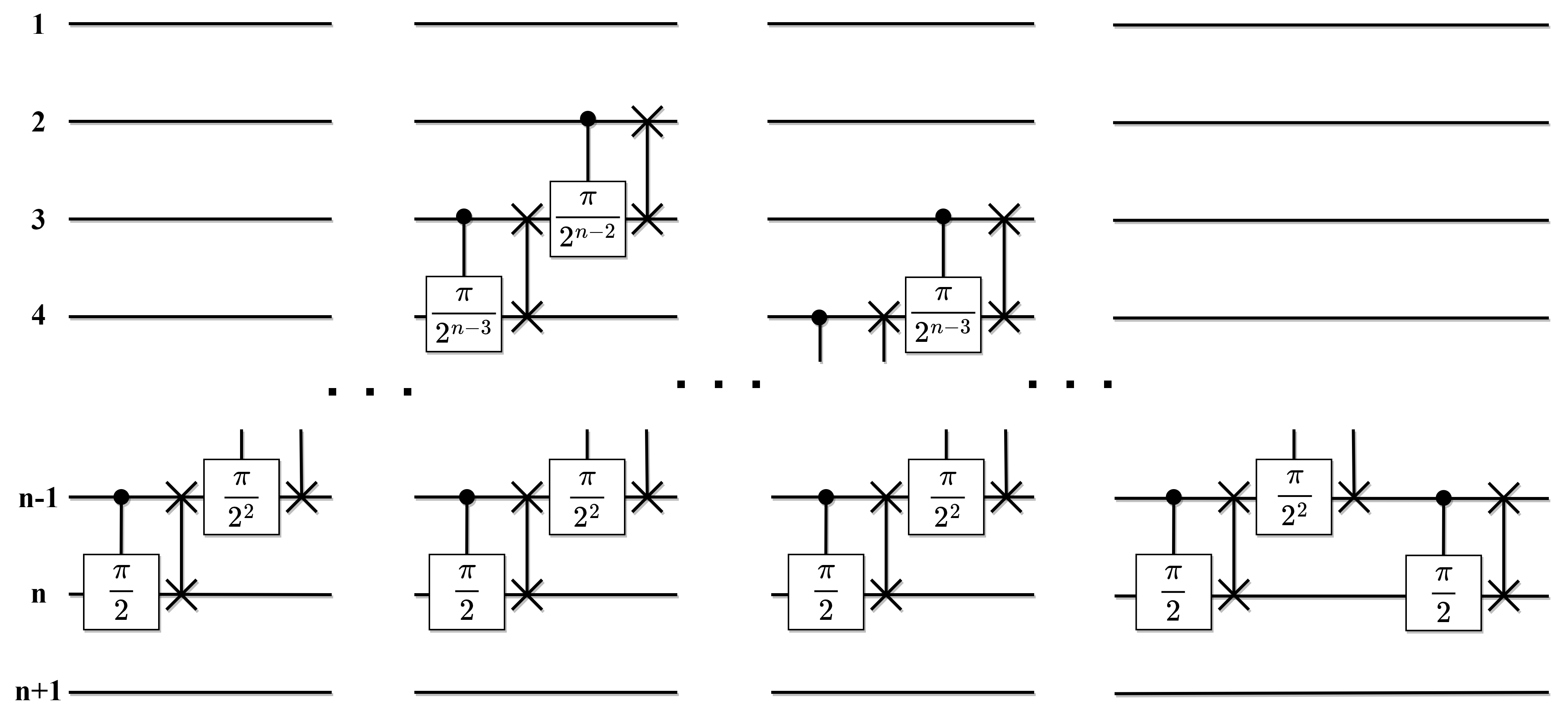}
		\end{minipage}
		\vspace{0.3cm}
		
		\leftline{~$C_3^\prime:~~d_{C_3^\prime}=4n-6$\hspace{6.15cm}$C_6^\prime:~~d_{C_6^\prime}=4n-10$}
		\begin{minipage}{0.49\linewidth}
			\vspace{0.1cm}
			\includegraphics[width=1\linewidth]{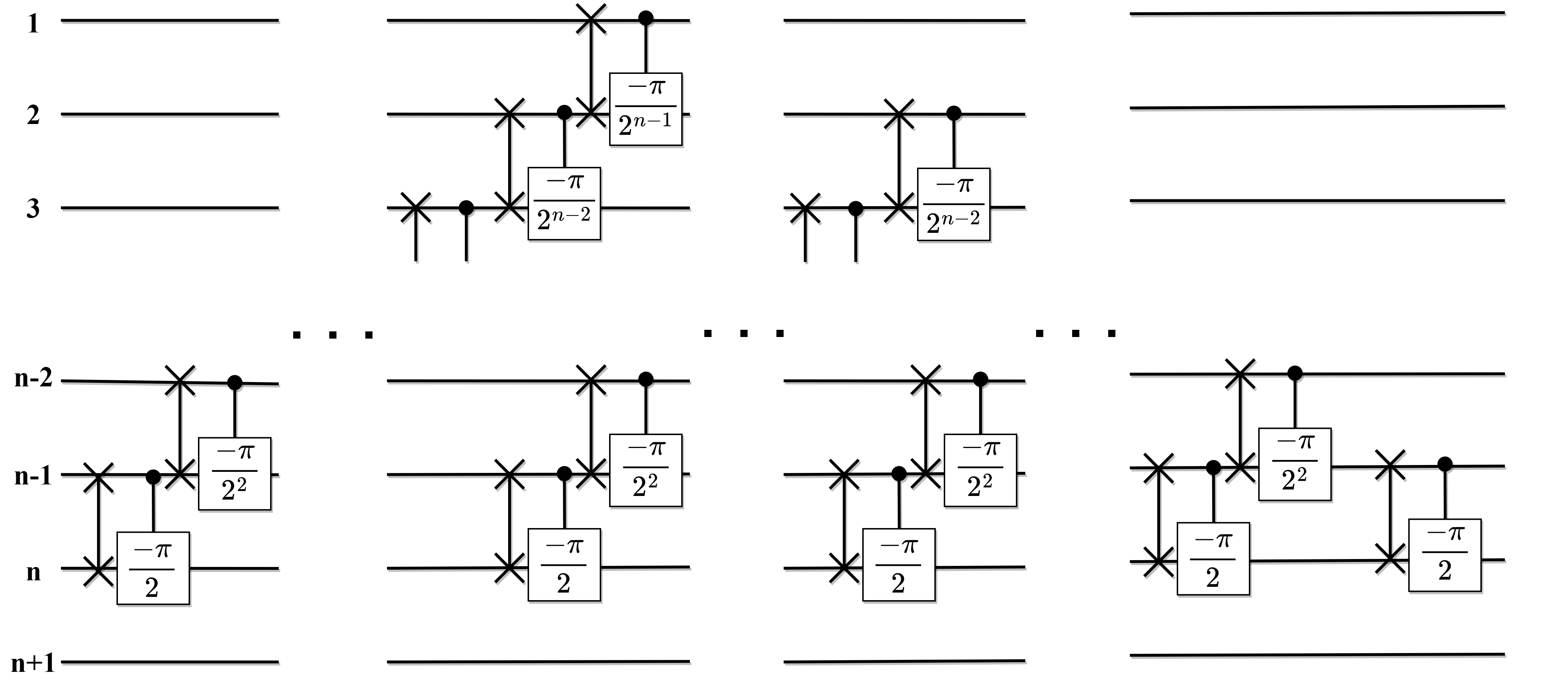}
		\end{minipage}
		\begin{minipage}{0.49\linewidth}
			\includegraphics[width=1\linewidth]{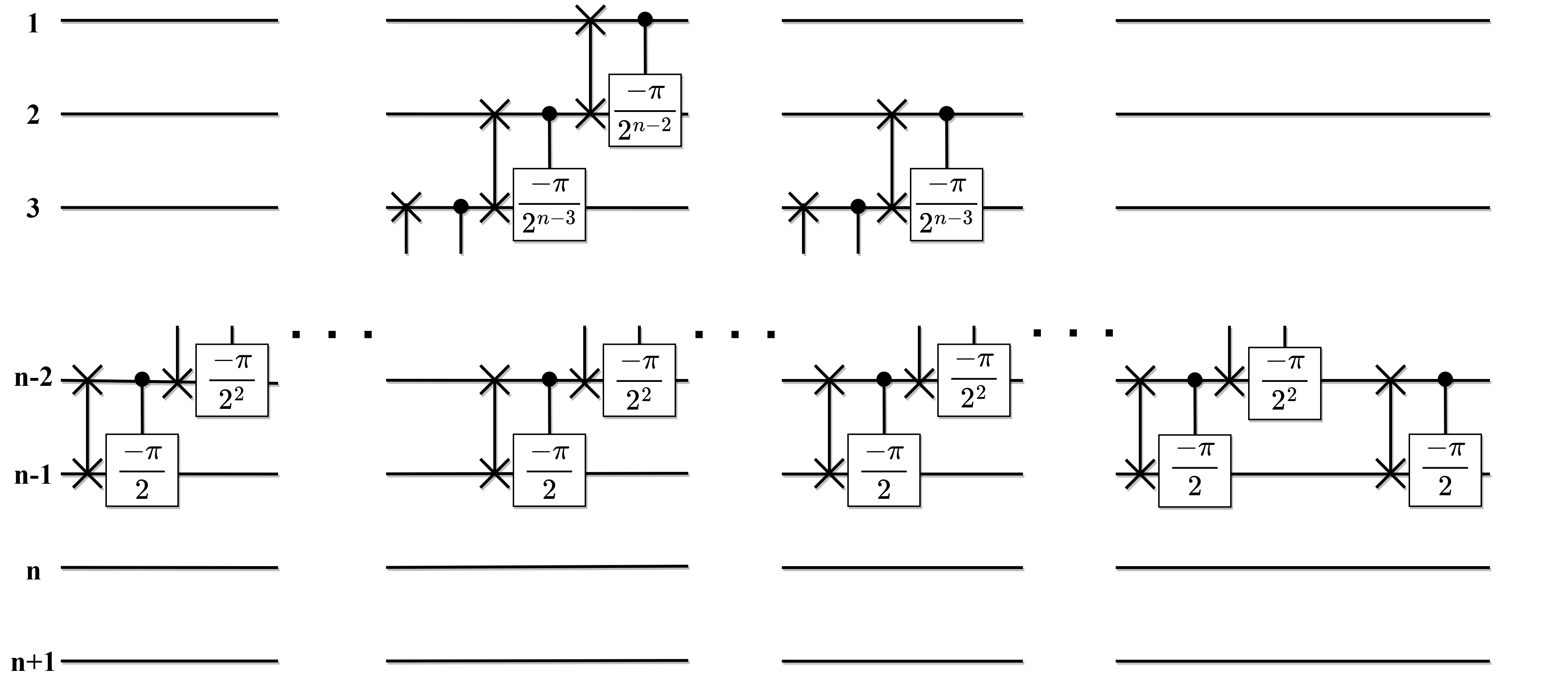}
		\end{minipage}
		\vspace{0.3cm}
	
		\leftline{~$C_2^\prime:~~d_{C_2^\prime}=n$\hspace{3.2cm}$C_5^\prime:~~d_{C_5^\prime}=n-1$\hspace{2.45cm}$C_{3.5}^\prime:~~d_{C_{3.5}^\prime}=n$\hspace{1.5cm}$C_{6.5}^\prime:~~d_{C_{6.5}^\prime}=n-1$}
		\begin{minipage}{0.29\linewidth}
			\vspace{0.2cm}
			\hspace{-0.4cm}\includegraphics[width=1\linewidth]{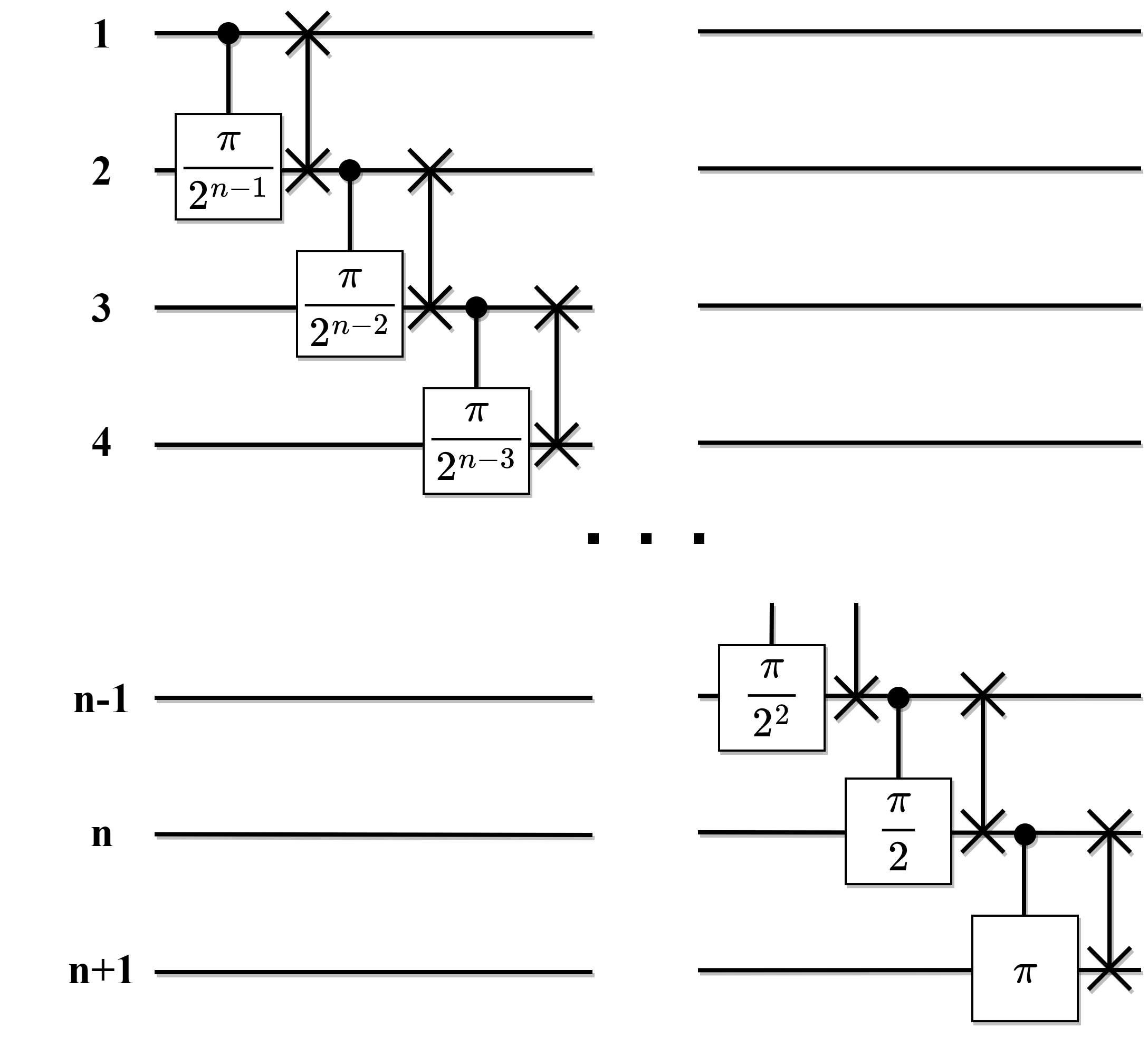}
		\end{minipage}
		\begin{minipage}{0.29\linewidth}
			\vspace{0.05cm}
			\hspace{-0.6cm}\includegraphics[width=1\linewidth]{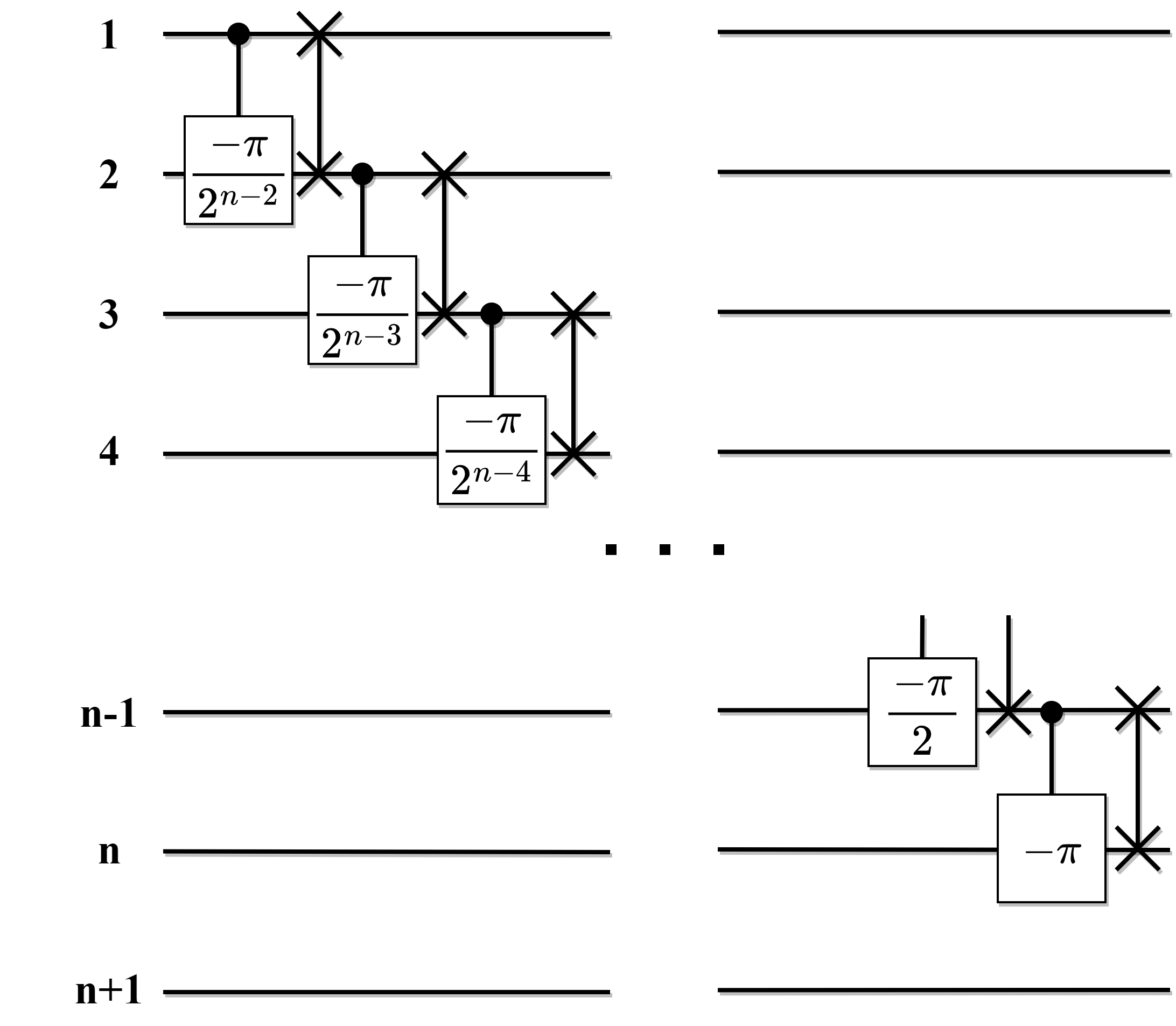}
		\end{minipage}
		\begin{minipage}{0.2\linewidth}
			\hspace{-0.3cm}\includegraphics[width=1\linewidth]{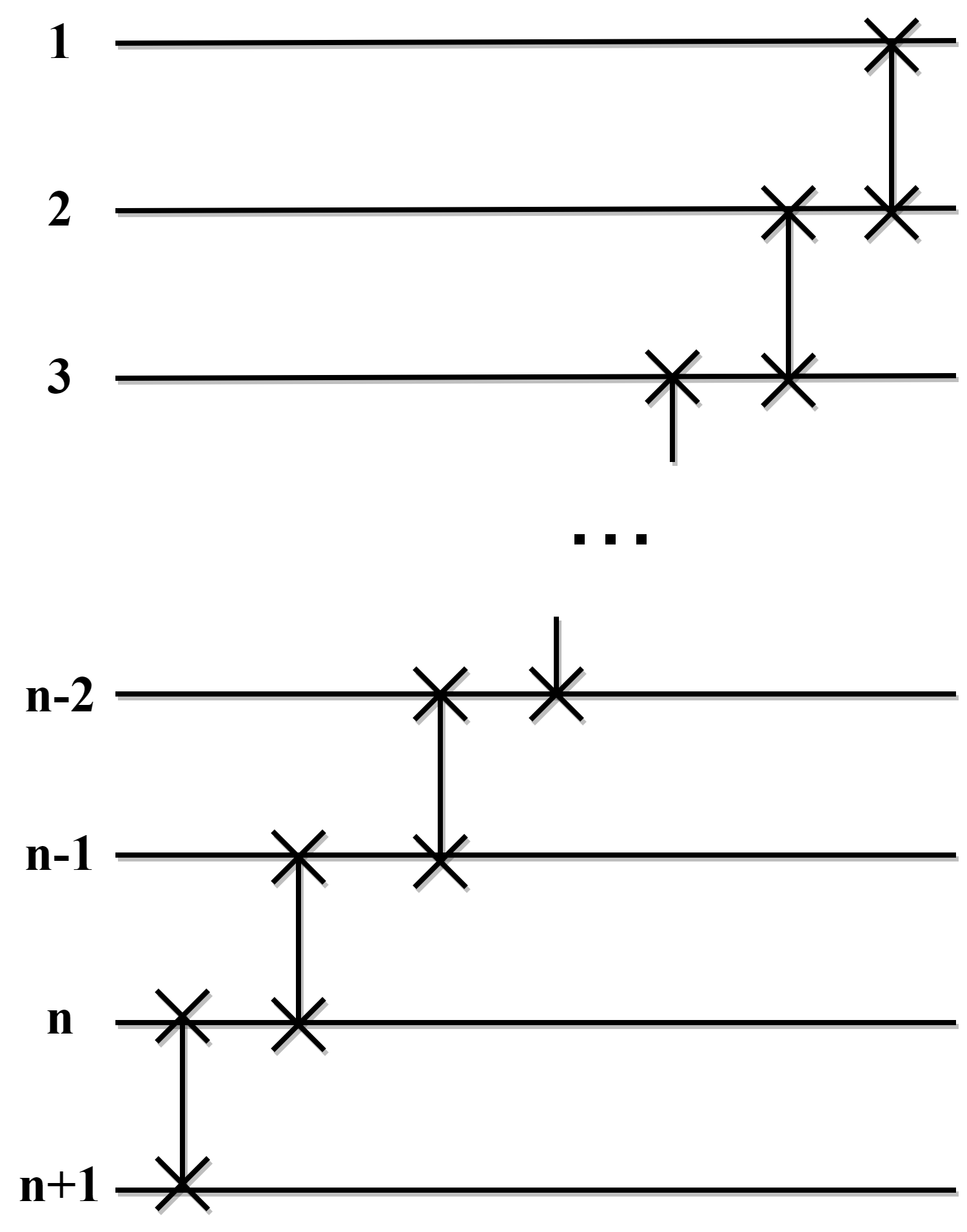}
		\end{minipage}
		\begin{minipage}{0.2\linewidth}
			\includegraphics[width=1\linewidth]{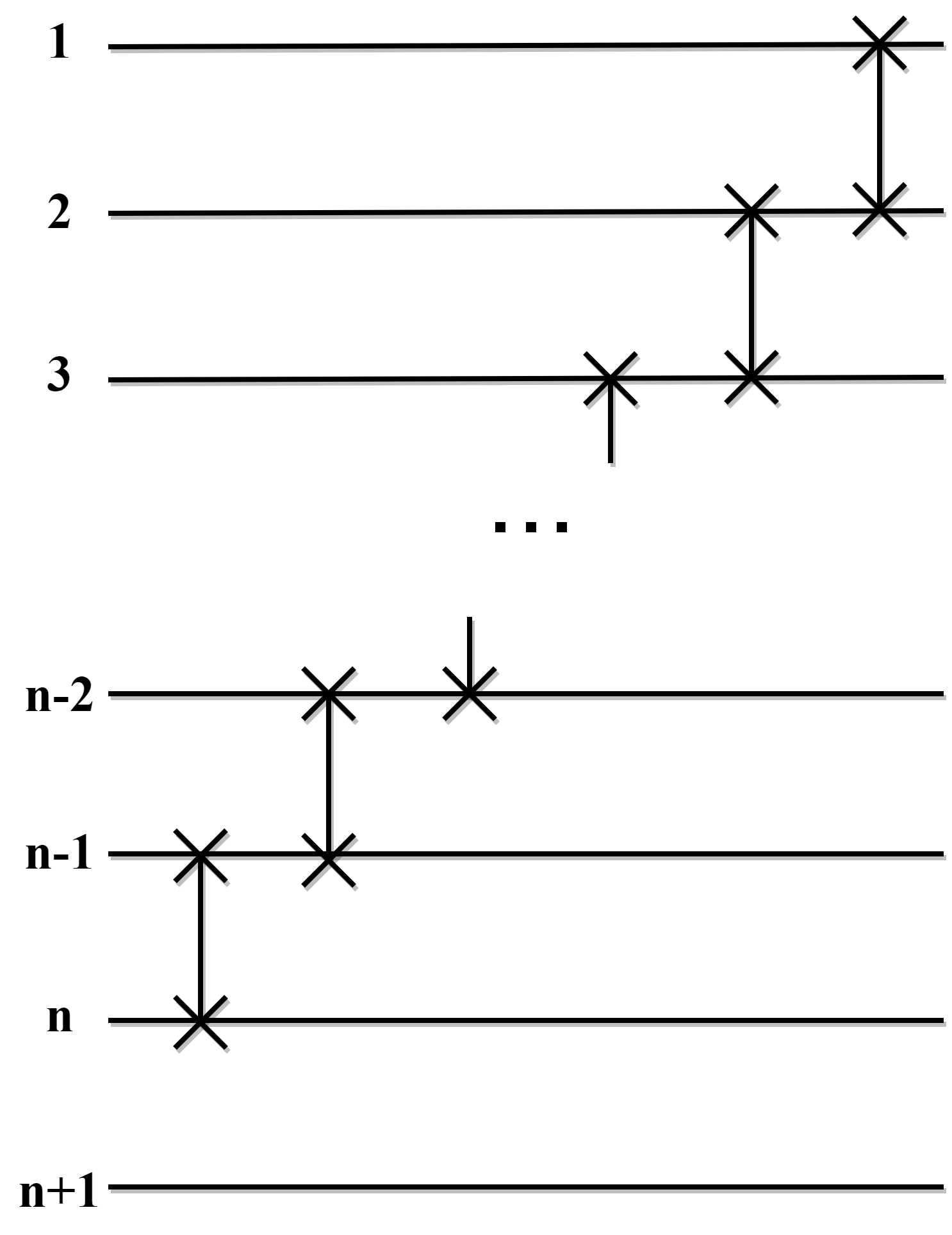}
		\end{minipage}
		\caption{\label{parts_local_gates}The decomposition for $C_1^\prime$, $C_2^\prime$, $C_3^\prime$, $C_{3.5}^\prime$, $C_4^\prime$, $C_5^\prime$, $C_6^\prime$ and $C_{6.5}^\prime$ with local gates set Eq. (\ref{Grover_local_gates_2}). They only include local controlled-$R_x(\theta)$ gates and local SWAP gates. $d$ is the circuit depth.}
	\end{figure*}

	We decompose $(n+1)$-qubit Toffoli gate as shown in Fig. (\ref{Toffoli_circuit}) and (\ref{parts_local_gates}) with the following local gates set \cite{leng2023modifying}:
	
	\begin{align}
		\{&[P_1]~\widetilde{S(\frac{\pi}{2^n})_1};~\widetilde{S(\frac{\pi}{2^n})_2}; ~\widetilde{S(\frac{\pi}{2^{n-1}})_3},~...,~\widetilde{S(\frac{\pi}{2^2})_n};\nonumber\\
		&[C_1^\prime]~\widetilde{\mathrm{C}R_x(\frac{\pi}{2^{n-1}})_{2,3}},~\widetilde{\mathrm{C}R_x(\frac{\pi}{2^{n-2}})_{3,4}},~...,~\widetilde{\mathrm{C}R_x(\frac{\pi}{2})_{n,n+1}};\nonumber\\
		&[C_2^\prime]~\widetilde{\mathrm{C}R_x(\frac{\pi}{2^{n-1}})_{1,2}},~\widetilde{\mathrm{C}R_x(\frac{\pi}{2^{n-2}})_{2,3}},~...,~\widetilde{\mathrm{C}R_x(\pi)_{n,n+1}};\nonumber\\
		&[C_3^\prime]~\widetilde{\mathrm{C}R_x(\frac{-\pi}{2^{n-1}})_{1,2}},~\widetilde{\mathrm{C}R_x(\frac{-\pi}{2^{n-2}})_{2,3}},~...,~\widetilde{\mathrm{C}R_x(\frac{-\pi}{2})_{n-1,n}};\nonumber\\
		&[P_2]~\widetilde{S(\frac{-\pi}{2^n})_2}, ~\widetilde{S(\frac{-\pi}{2^{n-1}})_3},~...,~\widetilde{S(\frac{-\pi}{2^2})_n};\nonumber\\
		&[C_4^\prime]~\widetilde{\mathrm{C}R_x(\frac{\pi}{2^{n-2}})_{2,3}},~\widetilde{\mathrm{C}R_x(\frac{\pi}{2^{n-3}})_{3,4}},~...,~\widetilde{\mathrm{C}R_x(\frac{\pi}{2})_{n-1,n}};\nonumber\\
		&[C_5^\prime]~\widetilde{\mathrm{C}R_x(\frac{-\pi}{2^{n-2}})_{1,2}},~\widetilde{\mathrm{C}R_x(\frac{-\pi}{2^{n-3}})_{2,3}},~...,~\widetilde{\mathrm{C}R_x(-\pi)_{n-1,n}};\nonumber\\
		&[C_6^\prime]~\widetilde{\mathrm{C}R_x(\frac{-\pi}{2^{n-2}})_{1,2}},\widetilde{\mathrm{C}R_x(\frac{-\pi}{2^{n-3}})_{2,3}},...,\widetilde{\mathrm{C}R_x(\frac{-\pi}{2})_{n-2,n-1}};\nonumber\\
		&[C_{1-6.5}^\prime]~\widetilde{\mathrm{SWAP}_{1,2}},~\widetilde{\mathrm{SWAP}_{2,3}},~...,~\widetilde{\mathrm{SWAP}_{n,n+1}}\}\label{Grover_local_gates_2}
	\end{align}
	where $S(\theta)$ is the $\theta$-phase gate $|0\rangle\langle0|+e^{i\theta}|1\rangle\langle1|$ and $\mathrm{C}R_x(\theta)$ is the controlled-$R_x(\theta)$ gate with $R_x(\theta)=e^{-iX\frac{\theta}{2}}$. Notations $[P_1],[C_1^\prime],...$ in Eq. (\ref{Grover_local_gates_2}) means that the decomposition of parts $P_1,C_1^\prime,...$ include the local gates in this line. For example, local gates in the first line are applied to construct $P_1$ and in the last line are applied to construct $C_1^\prime$, $C_2^\prime$, $C_3^\prime$, $C_{3.5}^\prime$, $C_4^\prime$, $C_5^\prime$, $C_6^\prime$ and $C_{6.5}^\prime$. The circuit depth under this decomposition is $d_{\mathrm{Toff}}=d_{P_1}+d_{C_1^\prime}+...+d_{C_{6.5}^\prime}=20n-32$.
	
	\section{Noise model}
	Suppose $G$ is an ideal unitary gate. The corresponding noisy quantum channel is defined as
	\begin{align}
		\mathcal{G}(\rho) = \mathcal{E}(p)(G\rho G^\dagger),
	\end{align}
	where
	\begin{align}
		\mathcal{E}(p)(\rho) = \sum_iE_i(p)\rho E_i(p)^\dagger,
	\end{align}
	and $p$ is the noise parameter.
	\subsubsection{Bit-flip noise}
	For a single-qubit gate $G_a$ applied on qubit $a$, the corresponding bit-flip channel is
	\begin{align}
		\mathcal{G}_a(\rho_a) = \mathcal{E}(p)(G_a\rho_a G^\dagger_a),
	\end{align}
	where
	\begin{align}
		E_1(p)=\sqrt{1-p}I,~~E_2(p)=\sqrt{p}X.
	\end{align}
	For a two-qubit gate $G_{ab}$ applied on qubit $a$ and $b$, we consider the following noisy channel \cite{endo2018practical}
	\begin{align}
		\mathcal{G}_{ab}(\rho_{ab}) = \mathcal{E}(2p)\otimes\mathcal{E}(2p)(G_{ab}\rho_{ab} G_{ab}^\dagger).
	\end{align}
	Idle qubit does not have bit-flip noise.
	
	\subsubsection{Phase-flip noise}
	For a single-qubit gate $G_a$, the corresponding phase-flip channel is
	\begin{align}
		\mathcal{G}_a(\rho_a) = \mathcal{E}(p)(G_a\rho_a G^\dagger_a),
	\end{align}
	where
	\begin{align}
		E_1(p)=\sqrt{1-p}I,~~E_2(p)=\sqrt{p}Z.
	\end{align}
	For two-qubit gate $G_{ab}$, we consider the following noisy channel
	\begin{align}
		\mathcal{G}_{ab}(\rho_{ab}) = \mathcal{E}(2p)\otimes\mathcal{E}(2p)(G_{ab}\rho_{ab} G_{ab}^\dagger).
	\end{align}
	Idle qubit does not have phase-flip noise.
	
	\subsubsection{Depolarizing noise}
	For a single-qubit gate $G_a$, the corresponding depolarizing channel is
	\begin{align}
		\mathcal{G}_a(\rho_a)=(1-\frac{4}{3}p)G_a\rho_a G^\dagger_a+\frac{4}{3}p\frac{\mathds{1}}{2} =\mathcal{E}(p)(G_a\rho_a G^\dagger_a),
	\end{align}
	where
	\begin{align}
		E_1(p)=&\sqrt{1-p}I,&E_2(p)=\sqrt{\frac{p}{3}}X,\nonumber\\
		E_3(p)=&\sqrt{\frac{p}{3}}Y,&E_4(p)=\sqrt{\frac{p}{3}}Z.
	\end{align}
	For two-qubit gate $G_{ab}$, we consider the following noisy channel
	\begin{align}
		\mathcal{G}_{ab}(\rho_{ab}) = \mathcal{E}(2p)\otimes\mathcal{E}(2p)(G_{ab}\rho_{ab} G_{ab}^\dagger).
	\end{align}
	Idle qubit does not have depolarizing noise.
	
	\subsubsection{Global depolarizing noise}
	Global depolarizing noise is defined as \cite{cao2022mitigating}
	\begin{align}
		\mathcal{E}_d(\rho_t)=p_d\frac{I}{2^t}+(1-p_d)\rho_t,
	\end{align}
	where $\rho_t$ is the state for all the $t$ qubits and subscript $d$ denotes the circuit depth. This noise occurs after we implement the whole quantum circuit $U$ \cite{cao2022mitigating}:
	\begin{align}
		\mathcal{U}(\rho_t)=\mathcal{E}_d(U\rho_tU^\dagger)=p_d\frac{I}{2^t}+(1-p_d)U\rho_tU^\dagger.\label{total_globel}
	\end{align}
	Here, we assume that the depolarizing probability $p_d$ depends on the circuit depth $d$ \cite{vrana2014fault,cohn2016grover}. Taking $d=1$, the circuit degenerates into a unitary operation $G^{(i)}$ (it is the direct product of some local gates). Then we obtain the corresponding noisy quantum channel
	\begin{align}
		\mathcal{G}^{(i)}(\rho_t)=&\mathcal{E}_1(G^{(i)}\rho_tG^{(i)\dagger})=p_1\frac{I}{2^t}+(1-p)G^{(i)}\rho_tG^{(i)\dagger}\nonumber\\
		=&G^{(i)}\mathcal{E}_1(\rho_t)G^{(i)\dagger}.
	\end{align}
	We see that $\mathcal{E}_1$ and $G^{(i)}$ commute with each other. A $d$-depth circuit $U=...G^{(k)}G^{(j)}G^{(i)}$ has $d$ unitary operations. Then we get the noisy quantum channel for $U$:
	\begin{align}
		\mathcal{U}(\rho_t)=&...\mathcal{G}^{(k)}\mathcal{G}^{(j)}\mathcal{G}^{(i)}(\rho_t)\nonumber\\
		=&...G^{(k)}G^{(j)}G^{(i)}\mathcal{E}_1^d(\rho_t)G^{(i)\dagger}G^{(j)\dagger}G^{(k)\dagger}...\nonumber\\
		=&U\left[\left(1-(1-p_1)^d\right)\frac{I}{2^t}+(1-p_1)^d\rho_t\right]U^\dagger\nonumber\\
		=&\left(1-(1-p_1)^d\right)\frac{I}{2^t}+(1-p_1)^dU\rho_tU^\dagger.
	\end{align}
	Comparing with Eq. (\ref{total_globel}), we obtain $p_d=\left(1-(1-p_1)^d\right)$. We have $(1-p_1)^d$ probability to obtain a correct state. Now consider the Grover circuit in Fig. \ref{Grover_circuit}. The circuit depth of $S_i$ is $d_{S_i}=2$. We have chosen $(n+1)$-qubit Toffoli gate as oracle operator $S_o$, so its circuit depth is $d_{S_o}=20n-32$ according to Appendix A. Then we have $d_{S_r}=20n-28$ and $d_{S_g}=d_{S_o}+d_{S_r}=40n-60$. The circuit depth of $k$-iterate Grover algorithm is $d_{U_k}=d_{S_i}+kd_{S_g}=40nk-60k+2$. So the success probability and running time of $k$-iterate Grover algorithm under time-independent global depolarizing noise is
	\begin{align}
		s_{\mathrm{doc}}(k)=&(1-p_1)^{d_{U_k}}s(k)+\left(1-(1-p_1)^{d_{U_k}}\right)\frac{1}{2^n},\nonumber\\
		r_{\mathrm{doc}}(k)=&\frac{k}{s_{\mathrm{doc}}(k)},
	\end{align}
	where $s(k)$ is given by Eq. (\ref{success_probability_noiseless}). We directly apply this analytical result for running time instead of calculating the Grover circuit. This allowed us to simulate for more qubits.
	
	\subsubsection{Cross-talk noise}
	Our noise-tolerant method is not suitable for cross-talk noise since it is not a time-independent noise. However, we will neglect this `mistake' and numerically simulate the Grover circuit under this noise. We apply a single-qubit gate $G_a$ or an identity gate $I_a$ on qubit $a$. If its nearby qubit is applied by a local gate, $G_a$ will be affected by bit-flip noise:
	\begin{align}
		\mathcal{G}_a(\rho_a) = \mathcal{E}(p)(G_a\rho_a G^\dagger_a),
	\end{align}
	where
	\begin{align}
		E_1(p)=\sqrt{1-p}I,~~E_2(p)=\sqrt{p}X.
	\end{align}
	Similar result arises for $I_a$. It means that cross-talk noise flip the nearby qubit with probability $p$. We apply a two-qubit gate $G_{ab}$ on qubit $a$ and $b$. If its nearby qubit is applied by a local gate, we consider the following noisy channel
	\begin{align}
		\mathcal{G}_{ab}(\rho_{ab}) = \mathcal{E}(p)\otimes\mathcal{E}(p)(G_{ab}\rho_{ab} G_{ab}^\dagger).
	\end{align}
	It means that cross-talk noise makes each of nearby two qubits flip with probability $p$. Consider one-dimensional qubit chain. The single-qubit gate $G_a$ or identity gate $I_a$ maybe sandwiched between two local gates. In this case, the corresponding noisy channel is
	\begin{align}
		\mathcal{G}^\prime_a(\rho_a)=\mathcal{E}^2(G_a\rho_a G^\dagger_a):=\mathcal{E}^\prime(G_a\rho_a G^\dagger_a),
	\end{align}
	where
	\begin{align}
		E^\prime_1(p)=\sqrt{1-2p+2p^2}I,~~E^\prime_2(p)=\sqrt{2p-2p^2}X.
	\end{align}
	Similar result arises for $I_a$. For two-qubit gate $G_{ab}$ sandwiched between two local gates, we consider the following noisy channel
	\begin{align}
		\mathcal{G}^\prime_{ab}(\rho_{ab}) = \mathcal{E}^\prime(p)\otimes\mathcal{E}^\prime(p)(G_{ab}\rho_{ab} G_{ab}^\dagger).
	\end{align}
	Although cross-talk noise is not time-independent, our noise-tolerant method still exponentially improves the noise threshold under this noise as shown in Fig. \ref{noise_threshold}. This suggests that our noise-tolerant method could be applied in more general noisy environment.

	\bibliography{refs}
	
\end{document}